\documentclass[pdftex,twocolumn,epjc3]{svjour3}         

\RequirePackage[T1]{fontenc}

\smartqed  

\RequirePackage{graphicx}
\RequirePackage{mathptmx}      
\RequirePackage{flushend}
\RequirePackage[numbers,sort&compress]{natbib}
\RequirePackage[colorlinks,citecolor=blue,urlcolor=blue,linkcolor=blue]{hyperref}
\RequirePackage{booktabs}

\journalname{Eur. Phys. J. C}

\usepackage[binary-units=true]{siunitx}
\DeclareSIUnit{\be}{BE}

\usepackage[modulo]{lineno}
\begin{document}


\title{A time resolved study of injection backgrounds during the first commissioning phase of SuperKEKB}

\author{Miroslav Gabriel\thanksref{addr1,e1}
       \and
       Frank Simon\thanksref{addr1,e2}
       \and 
       Hendrik Windel\thanksref{addr1} 
       \and
       Yoshihiro Funakoshi\thanksref{addr2}
       \and
       Michael Hedges\thanksref{addr3, n1}
       \and
       Naoko Iida\thanksref{addr2} 
       \and
       Igal Jaegle\thanksref{addr3, n2}
       \and  
       Christian Kiesling\thanksref{addr1}
       \and
       Naomi van der Kolk\thanksref{addr1,n3} 
       \and
       Peter Lewis\thanksref{addr3,n4}
       \and 
       Hiroyuki Nakayama\thanksref{addr2}
       \and
       Yukiyoshi Ohnishi\thanksref{addr2}
       \and
       Riccardo de Sangro\thanksref{addr4}
       \and
       Yusuke Suetsugu\thanksref{addr2}
       \and
       Marco Szalay\thanksref{addr1,n5} 
       \and
       Sven Vahsen\thanksref{addr3}       
}

\thankstext{e1}{e-mail: mgabriel@mpp.mpg.de}
\thankstext{e2}{e-mail: fsimon@mpp.mpg.de}
\thankstext{n1}{now at Purdue University, West Lafayette, IN, 47907, U.S.A.}
\thankstext{n2}{now at Thomas Jefferson National Accelerator Facility, Newport News, VA, 23606, U.S.A.}
\thankstext{n3}{now at The Hague University of Applied Sciences, The Netherlands.}
\thankstext{n4}{now at University of Bonn, 53115 Bonn, Germany}
\thankstext{n5}{now at Google LLC., Santa Barbara, CA, 93101, U.S.A.}

\institute{Max-Planck-Institut f\"ur Physik, 80805 M\"unchen, Germany\label{addr1}
\and
High Energy Accelerator Research Organization (KEK), Tsukuba 305-0801, Japan\label{addr2}
\and
University of Hawaii, Honolulu, Hawaii 96822, U.S.A.\label{addr3}
\and
INFN Laboratori Nazionali di Frascati, I-00044 Frascati, Italy\label{addr4}}

\date{October 13, 2021}
%
%
\maketitle
%

\begin{abstract}

We report on measurements of beam backgrounds during the first commissioning phase of the SuperKEKB collider in 2016, performed with the plastic scintillator and silicon photomultiplier-based CLAWS detector system. The sub-nano\-second time resolution and single particle detection capability of the sensors allow bunch-by-bunch measurements, enable CLAWS to perform a novel time resolved analysis of beam backgrounds, and make the system uniquely suited for the study of injection backgrounds.
We present measurements of various aspects of regular beam background and injection backgrounds which include time structure and decay behavior of injection backgrounds, hit-energy spectra and overall background rates.
These measurements show that the elevated background rates following an injection generally last for several milliseconds, with the majority of the background particles typically observed within the first \SI{500}{\micro\second}. The injection backgrounds exhibit pronounced patterns in time, connected to betatron and synchrotron oscillations in the accelerator rings. The frequencies of these patterns are determined from detector data. 
\end{abstract}
%

\section{Introduction}
\label{sec:Intro}

The SuperKEKB accelerator~\cite{Akai:2018mbz} is a next generation flavor factory currently operating in Tsukuba, Japan.
It is an asymmetric-energy electron-positron collider which runs predominantly at a center-of-mass energy corresponding to the mass of the $\Upsilon (4S)$ resonance of \SI{10.58}{\GeV}. The beam energies are \SI{7}{\GeV} for electrons in the high energy ring (HER) and  \SI{4}{\GeV} for positrons in the low energy ring (LER).
Compared to its predecessor KEKB, SuperKEKB has a 40 times higher design luminosity of \SI[exponent-product = \times]{8e35}{\centi  \meter ^{-2}  \second ^{-1}}.
This gain in luminosity is mainly achieved by changes in two operational parameters: doubling of the beam currents to \SI{3.2}{\A} (HER) and \SI{2.6}{\A} (LER), and extreme focusing of the beams at the interaction point (IP) to vertical beam sizes of \SI{50}{\nano\meter} using the so-called nano-beam scheme~\cite{Bona:2007qt}. 
Despite the short beam lifetimes on the order of minutes, high integrated luminosities are achieved by continuously injecting new particles at a rate of up to \SI{25}{Hz} per beam.
 Both rings have a circumference of \SI{3016.3}{\meter} which results in a revolution time, $T_{rev}$, of \SI{10061.4}{\ns}.

The ambitious luminosity goals are expected to cause challenging levels of beam backgrounds in various subsystems of the corresponding Belle~II experiment~\cite{Abe:2010gxa,Lewis:2018ayu}.
In particular the backgrounds related to the continuous top-up injections of new beam particles, or \textit{injection backgrounds}, may lead to background levels which require an interruption of data recording in the detector during the transit of injection bunches. Thus, a detailed understanding of the backgrounds is of high importance in order to limit their impact on the Belle~II physics program.

To prime SuperKEKB and Belle~II for physics operation and full luminosities, a three-phase commissioning campaign has been performed.
Phase 1 focused on the basic operation of the accelerator and thus the final focusing system and the detector had not yet been installed at the IP and no intended collisions took place.
The primary objectives were improving vacuum conditions by scrubbing and performing basic machine studies such as tuning of the feedback system or low emittance optimization of the beam optics.
Focused beams and collisions always cause luminosity dependent backgrounds which make up a significant part of the overall backgrounds.
Measurements performed during Phase 1, by contrast, represented the unique opportunity to study beam-induced backgrounds and injection backgrounds independently in a collision-free environment. For this purpose, a set of dedicated beam background detectors was installed around the IP, collectively referred to as BEAST~II. Details on this system, and results of a first comprehensive analysis of the beam backgrounds at the IP have already been reported in Ref.~\cite{Lewis:2018ayu}.

One of the subsystems of BEAST~II is the CLAWS (s\textbf{C}in\-til\-lat\-ing \textbf{L}ight \textbf{A}nd \textbf{W}aveform \textbf{S}ensors) detector system \cite{Gabriel2019}. It consists of eight plastic scintillators with directly coupled silicon photomultipliers (SiPM) read out by commercial electronics and a custom-made DAQ
capable of continuously re\-cord\-ing data over periods up to several milliseconds. Its primary goal is a novel time resolved analysis of backgrounds, focussing on injection backgrounds in particular. The CLAWS system is uniquely capable of a continuous monitoring of the background conditions following top-up injections via the detection of charged particles, complementing the capabilities of other detector systems in BEAST~II. On February 10, 2016 and on February 26, 2016, CLAWS observed the very first beam bunches which were successfully circulated in the LER and the HER of SuperKEKB, respectively~\cite{firstbeams}. It should be noted that the beam conditions during Phase 1 differ substantially from those during later physics operations. The absence of the final focusing elements results in vertical beam sizes approximately three orders of magnitude larger than the final design values, and the maximum beam currents were a factor three below the design values in both rings. The positrons are created from the primary electron beam in a production target and thus have a significantly higher emittance at the source. The damping ring which reduces this emittance prior to injection was not yet operational during Phase 1. This resulted in increased background levels in the LER, as discussed throughout this paper. Due to these differences between Phase 1 and later phases, only limited quantitative information for the conditions at Belle II can be extracted. Nevertheless, the studies during Phase~1 contribute significantly to the overall understanding of the behaviour of backgrounds in SuperKEKB.

In this publication we report on measurements performed by CLAWS during Phase 1 of the SuperKEKB commissioning campaign. The present analysis updates and extends the CLAWS results included in~\cite{Lewis:2018ayu}. Section~\ref{sec:backgrounds} briefly introduces relevant sources of beam backgrounds at SuperKEKB. In Section~\ref{sec:setup}, we give an overview of the experimental setup of CLAWS and its data processing and event selection. Section~\ref{sec:time_structure} begins the discussion of the results by presenting the time structure of beam backgrounds recorded during periods with continuous top-up injections. Here, we also describe a number of different timing patterns encountered in the data and discuss their connection to betatron and synchrotron oscillations performed by the beam particles. Subsequently, Section~\ref{sec:hit_energy} studies the composition of backgrounds by examining the distributions of hit energies and Section~\ref{sec:decay_behavior} demonstrates a method for quantifying deposited energies and decay behavior of injection backgrounds with respect to accelerator conditions. Building on this, Section~\ref{sec:timing} presents a detailed analysis of the observed timing patterns. Finally, we briefly summarize the key findings in Section~\ref{sec:summary}.


\section{Background sources}
\label{sec:backgrounds}
The term \textit{backgrounds} generally refers to undesirable particles and can be divided into three types of processes: backgrounds caused by the circulating beams, or \textit{beam-induced backgrounds}; backgrounds as a byproduct of collisions, also referred to as \textit{luminosity-dependent backgrounds}; and backgrounds due to the injection of new beam particles, or \textit{injection backgrounds}. These background particles can be the beam particles themselves or secondary particles generated by beam particles which are colliding with accelerator elements.

Due to the lack of collisions, only beam-induced and injection backgrounds are relevant for the study of Phase~1 data. In the following, we provide an overview of the four main processes for these backgrounds.

\paragraph{Synchrotron radiation}
The first source of backgrounds is synchrotron radiation (SR), i.e. photons which are emitted by the beam particles while undergoing centripetal acceleration in the bending and focusing magnets. For Phase~1, simulations predict a moderate soft x-ray photon rate with an upper limit of \SI{500}{\Hz}. SR photons are therefore assumed to have no relevance~\cite{Lewis:2018ayu}.

\paragraph{Touschek scattering}
The term Touschek effect, or Touschek scattering, refers to an intra-bunch scattering process in which a single Coulomb scattering between two particles of the same bunch changes their energy significantly \cite{Piwinski:1998qs}.
This causes beam particles to depart from their nominal orbits, leading to significantly increased
betatron oscillations and consequently a larger emittance. When these particles exceed the acceptance of the accelerator, they collide with the beam\-pipe or other parts of the accelerator infrastructure and cause secondary particle showers. The higher beam currents and the dramatically reduced beam sizes due to the nano-beam scheme significantly increases the particle loss rates by Touschek scattering compared to the situation at KEKB.

\paragraph{Beam-gas scattering}
The third source of beam backgrounds is beam-gas scattering, which denotes scattering of beam particles with residual gas molecules in the beam pipe. Such an interaction can take place through two different processes: elastic Coulomb scattering and inelastic Bremsstrahlung, resulting in scattering of beam particles and photon production, which in turn can lead to secondary particles.

\paragraph{Injection background}
The final source of backgrounds considered here are injection backgrounds. SuperKEKB uses \textit{top-up injections} to compensate for particle losses due to background processes and collisions while maintaining constant luminosities. In such top-up injections, new bunches of electrons and positrons are continuously injected into already circulating bunches at a rate of \SI{25}{Hz} per ring. According to Liouville's theorem, the injected particles can not be incorporated into the same phase space volume as the circulating beams.
SuperKEKB therefore uses the so-called \textit{betatron injection scheme}~\cite{Mori:2014lda}, in which new particles are injected with deliberately large betatron oscillation amplitudes in the accelerator plane, which initially separates them from the circulating beams. As a consequence, the injected particles subsequently perform betatron oscillations around the closed orbit. Over the course of several hundred turns in the accelerator, these oscillations are then steadily reduced by synchrotron radiation damping until the injected particles have fully merged into the already circulating bunches. 

Beam particles for which these betatron oscillation amplitudes exceed the acceptance of the accelerator collide with the beampipe or other accelerator elements and lead to beam backgrounds. The cooling phase following after the injection of new particles can thus result in significantly increased background levels for the detectors whenever the respective particle bunches pass by the IP. These bunches are therefore referred to as \textit{injection bunches} or \textit{noisy bunches}. The higher emittance of the positrons compared to the electrons (see Section \ref{sec:Intro}) is expected to result in increased injection backgrounds in the LER compared to those in the HER.

A more elaborate description of the different background processes can be found in~\cite{Lewis:2018ayu}. Since the present paper focusses on injection backgrounds, the other background sources are usually not further differentiated in the discussions, and are referred to with the term {\it regular backgrounds} for simplicity.


\section{Experimental setup, data reconstruction and event selection}
\label{sec:setup}

A time-resolved analysis of injection backgrounds requires a detector that is capable of detecting signals as low as single minimum-ionizing particles. In order to differentiate between energy deposits of individuals bunches, a time resolution smaller than the time between consecutive bunches ($T_{bunch}=\SI{3.930}{\nano\second}$) is needed. Capturing the full time evolution of injection backgrounds requires to monitor background rates over hundreds of turns in the accelerator or up to several milliseconds. In this section, an overview of the setup, the data processing and the event selection of the CLAWS detector is provided. 


\subsection{Experimental setup}
\label{sec:setup:setup}
\begin{figure} 
 	\centering
 	\includegraphics[width=0.9\columnwidth]{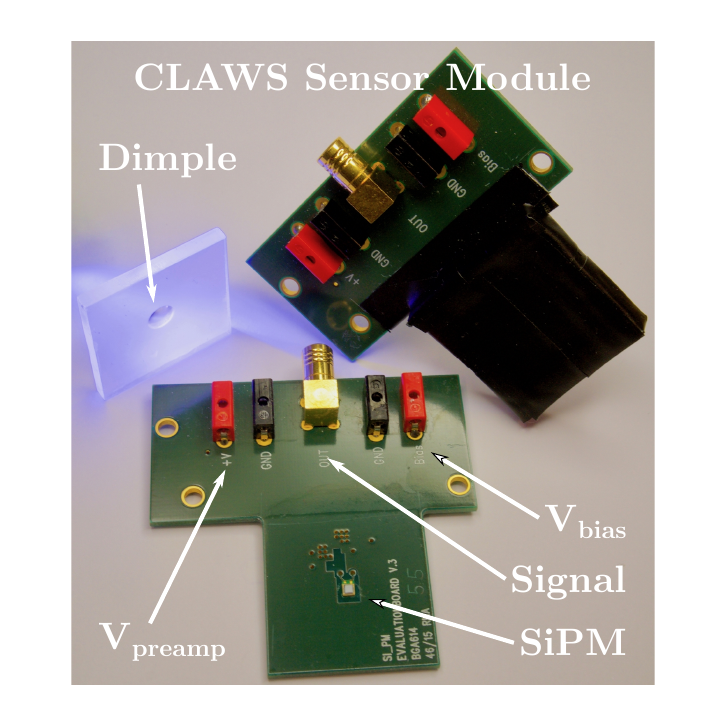}
 \caption{Photograph of an assembled CLAWS sensor module and its main components: a \SI{30 x 30 x 3 }{\milli\meter} scintillator tile and the PCB with photon sensor, preamplifier and connectors for power supply and signal transmission.}
 \label{fig:setup:sensors}
\end{figure}

%
The CLAWS sensors use plastic scintillator tiles read out with silicon photomultipliers (SiPMs) for the detection of the background particles. Figure \ref{fig:setup:sensors} shows an assembled sensor, as well as a bare scintillator tile and the readout unit with photon sensor.

The scintillator material has a polystyrene base (Styron 143E) with additional small admixtures of scintillating fluors (p-terphenyl, 2\%) and POPOP (0.05\%), and emits photons in the blue visible range with a peak emission of around \SI{420}{\nano\meter}~\cite{Hahn:1404985}. The design of the scintillator tiles is based on the SiPM-on-tile technology developed for highly granular analog hadronic calorimeter of the CALICE collaboration \cite{Sefkow:2018rhp}. The tiles are cuboid-shaped with a size of \SI[product-units = power]{30 x 30 x 3}{\milli\meter} and have a spherical dimple drilled in the center of the bottom face of the tile at the position of the photon sensor \cite{Simon:2010hf,Liu:2015cpe}. The dimple has a depth of \SI{1.5}{\milli\meter} and diameter on surface of \SI{6}{\milli\meter}.

The scintillation photons are captured by SMD-type Ha\-ma\-matsu MPPC S13360-1325PE series SiPMs~\cite{mppc:2019}, which have 2668 pixels with a size of \SI[product-units = power]{25 x 25}{\micro\meter} distributed over an active area of \SI[product-units = power]{1.3 x 1.3}{\milli\meter}, a thermal dark rate (at room temperature) as low as \SI{70}{\kilo\hertz} and a crosstalk probability of 1\%.
Together, the scintillating tiles and the SiPMs are mounted on electronics boards that house a pre-amplifier and provide signal transmission, power distribution and mechanical support. The design of these sensor modules is based on the CALICE-T3B experiment \cite{Simon:2013zya, Adloff:2014rya} which was used for the study of the time structure of hadronic showers in highly granular calorimeters. The detectors are primarily sensitive to charged particles, but in principle also show responses to high-energy photons and to MeV neutrons.

\begin{figure*}
\centering
\includegraphics[width=0.55\textwidth]{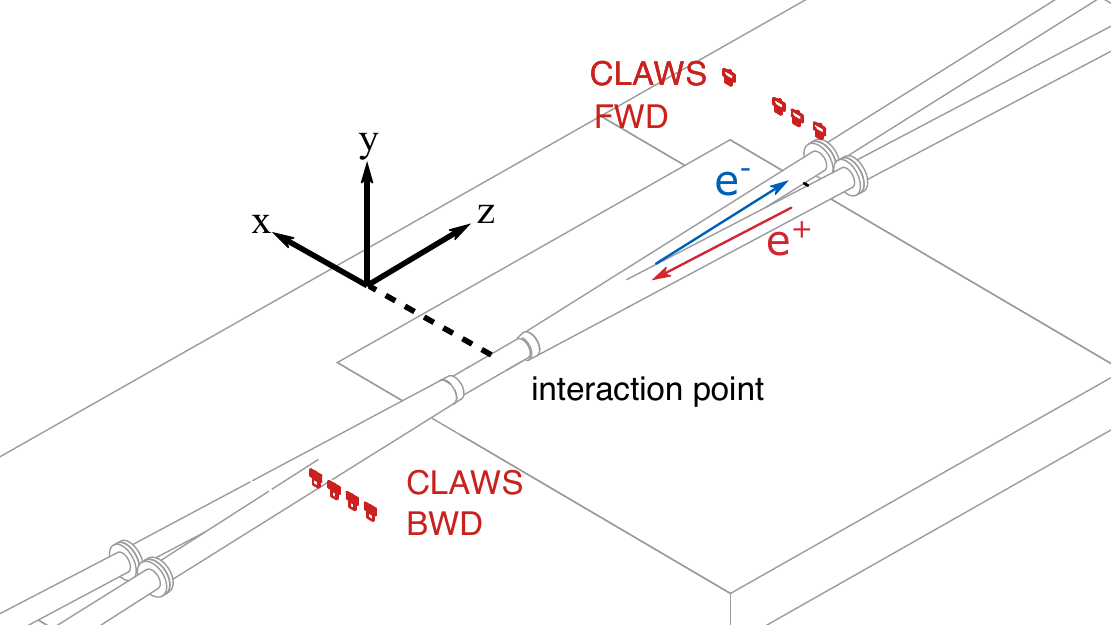}
\hfill
\includegraphics[width=0.445\textwidth]{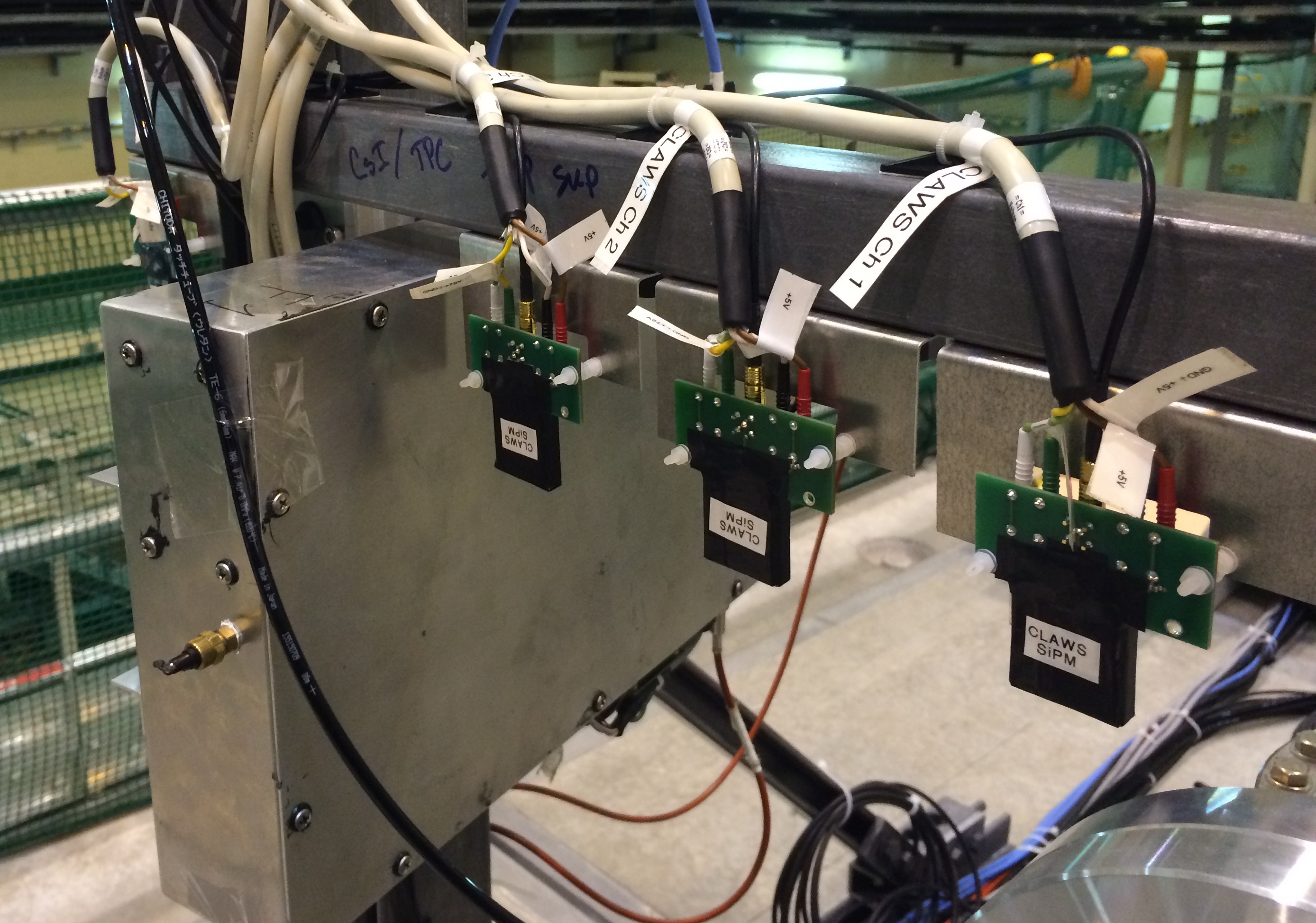}
\caption{{\it Left:} Layout of the CLAWS installation around the SuperKEKB interaction point. The positrons in the LER travel from top right to bottom left, the electrons in the HER from bottom left to top right. {\it Right:} Photograph of the CLAWS FWD station, showing the three inner sensors FWD1-3. The FWD4 sensor is not fully visible, as it is located behind another BEAST II detector element. The beampipe is visible in the lower right corner of the image.}
\label{fig:CLAWSSetup}
\end{figure*}

To cover the full range of expected rates, eight CLAWS sensor modules were installed at different distances from the beams, arranged in two stations with four detectors each, as illustrated in Figure \ref{fig:CLAWSSetup}. In both stations, the modules are mounted directly on the BEAST~II support structure~\cite{Lewis:2018ayu} such that they are positioned in the plane of the accelerator in a line roughly perpendicular to the beam with a spacing of around \SI{10}{\centi\meter} between each detector. The sensors in the forward station (referred to as \emph{FWD1 - 4}) are located outside of the accelerator ring, next to the HER and downstream of the IP (with respect to the HER); the region for which simulations predicted the highest background rates.  The detectors of the backward station (referred to as \emph{BWD1 - 4}) are placed inside of the ring, again next to the HER but upstream of the IP, the region with the lowest background predictions. The sensors are numbered such that their distance from the beam increases with increasing number, with FWD1 / BWD1 closest to, and FWD4/ BWD4 furthest from the beam, respectively. During the runtime of Phase 1, measurements from all sensors provided data for an online monitor used to tune accelerator parameters and to commission SuperKEKB. For the presented analysis only data from the channels FWD1-3 are used, for which the full analog waveform data was saved throughout Phase 1. Despite the location of the FWD sensors on the outside of the LER, they are also sensitive to particles originating from the HER at positions further upstream, allowing studies of background properties in both rings.

The exposure to ionizing particles throughout beam operations leads to  radiation damage of the SiPMs and of the other detector components. A significant increase of the dark rate of the photon sensors has been observed, varying with the distance from the beam and depending on the detector location. The most exposed sensor was BWD1, followed by FWD1. No detailed dose measurements for the sensor position exist. Based on dosimeter information in nearby regions available for parts of the data taking period, and based on the observed particle rates, the TID received by these sensors was estimated to several 10 Gy. This resulted in an increase of the dark count rate by more than two orders of magnitude, beyond several 10 MHz. The BWD1 and FWD1 sensors were replaced after half of the run time to limit the negative effect of the increased dark rate on the waveform reconstruction.

The analog signals of the sensors are digitized with two 4-channel USB-controlled Picotech PicoScope 6404D oscilloscopes which sample each channel with 8 bit resolution at 1.25 GHz~\cite{pico6404d:2018}.
Each oscilloscope has a shared internal buffer of  $2\times 10^9$ samples, which allows the recording continuous waveforms of up to \SI{400}{\milli\second} in length for all four channels simultaneously. This enables the uninterrupted monitoring of particle rates over up to around 40000 consecutive turns in SuperKEKB. 

The oscilloscopes, as well as a power supply unit and a data acquisition and control computer, are located in a dedicated counting room outside of the radiation area. The analog signals of the sensors are transmitted via coaxial cables over a distance of 40 meters from the detector location to the counting room. To ensure sufficient signal quality, an additional analog amplification stage is included in the signal path at a distance of \SI{3}{m} from the detector location.

The CLAWS system uses two different types of trigger for regular data taking. For the study of injection backgrounds, the trigger pulse sent by the SuperKEKB injector linac prior to the injection of particles into the accelerator rings is used to trigger the system. For the measurement of non-injection backgrounds a random trigger is used, which is implemented as a pre-defined waiting time after the completion of a readout cycle.
The time required to read out and process the waveforms captured following a single trigger signal is several tens of seconds. CLAWS is thus not able to record at the injection rate of \SI{25}{Hz} and only observed a very small fraction of the injection events.

%

\subsection{Data calibration and processing}

Over the course of Phase 1, the CLAWS experiment recorded a total of 5 TB of raw data. This raw data primarily consists of waveforms capturing the output of the sensors. We differentiate between two different types of such raw waveforms: \textit{calibration waveforms} and \textit{physics waveforms}. A set of one physics waveform per channel recorded simultaneously following a common trigger is referred to as an \textit{event}. A \textit{run} consists of multiple events ranging from single events in special cases up to a few hundred events.

The calibration waveforms capture the pulses of single pixels firing due to thermal excitation (dark noise), and are used to obtain the average response to a single photon detected by the SiPMs. This average response incorporates all temperature effects and the majority of the readout effects. For each channel, 1000 calibration waveforms are recorded at the end of each run. This  updates the calibration used for the interpretation of the physics waveforms approximately every 30 minutes, which automatically corrects for variations in environmental and operational parameters. Due to the stable temperature conditions in the BEAST enclosure during running, the run-by-run variations of the average single p.e. amplitude are generally smaller than 1\%.

For the physics waveforms, the oscilloscopes continuously sample the sensor output over \SI{2.4}{\milli\second} during regular operation. Following the technique developed in~\cite{Simon:2013zya}, the processing of this raw data is built around the so called \textit{waveform decomposition}, in which the average response of a single photon is iteratively subtracted from a raw physics waveform in order to retrieve the time distribution of the scintillation photons. The single photon response is obtained from the calibration waveforms updated once per run. This process provides the arrival times of individual detected photons in bins of \SI{0.8}{\ns}, converting the analog signal to a time distribution of photons.

For each sensor module, the response to a traversing minimum-ionizing particle was established in additional calibration measurements performed in a climate controlled laboratory environment after the run time of Phase 1. Several thousand approximately perpendicularly incident cosmic muons were selected by a coincidence trigger formed by two additional sensor modules placed above and below the sensor under study. The most probable number of recorded scintillation photons is used to normalize the distribution of photons, i.e. convert it into units of MIPs. For all sensors, the most probable number of scintillation photons as a result of a traversing cosmic muon is between 13 and 16.5. Since this calibration was performed with the sensors having collected their full radiation dose, the calibration factors for the most irradiated sensors FWD1 and BWD1 may be influenced by the increased dark rate, resulting in an overestimation of the true response to minimum-ionising particles.

Also after the radiation damage collected during operations in Phase 1, the rate of having three or more simultaneous photons due to thermal dark rate does not exceed \SI{1}{\kilo\hertz}~\cite{Gabriel2019}, resulting in negligible background in terms of noise hits wrongly identified as an ionising particle. Similarly, imperfectly shielded ambient light is excluded as a source of observed signals since pedestal runs, taken at the end of the runtime, do not show a significant number of hits. The expected number of p.e.\ due to dark noise within the integration window of a MIP signal is expected to be below 3 also at the highest dark count rates encountered in Phase 1, and thus stays below 20\% of the MIP most probable value.

\subsection{Data set and event selection}
\label{sec:setup:datasets}

For the present analysis, all fully processed runs recorded between May 15 and June 28, 2016 are considered.\footnote{Events containing all information necessary for a full time resolved analysis have only been taken from May 15, 2016 onwards.} For these runs, events are classified based on the beam current in the HER, $I_{HER}$, the current in the LER, $I_{LER}$, and the type of injection to generate 15 distinct data sets that reflect a variety of different accelerator conditions. The data sets are summarized in Table~\ref{tab:setup:datasets}. Due to the long readout time of the system of several tens of seconds (see Section \ref{sec:setup:setup}), the data sets do not contain consecutive injection triggers. The larger data sets with several 1000 events combine data recorded over several days, up to the full period of data taking.

\begin{table}%
\centering
	\caption{Summary of data sets examined for the time resolved analysis of backgrounds. Beam currents are stated by the range between the minimum and maximum of currents in all events in the data set.}

\begin{tabular}{@{}lcccr@{}}
\toprule
Data Set & $BG$ & $I_{HER}$ & $I_{LER}$ & $N_{evts}$\\
 & & [\si{\mA}] & [\si{\mA}] & [\#] \\
\midrule
NI-ALL & - & \multicolumn{1}{r}{29 - 788} & \multicolumn{1}{r}{9 - 852} & 2676\\
NI-HER & - & \multicolumn{1}{r}{38 - 705} & 0 & 1806\\
NI-LER & - & 0 & \multicolumn{1}{r}{4 - 974} & 4057\\
NI-VACS & - & \multicolumn{1}{r}{452 - 706} & \multicolumn{1}{r}{105 - 752} & 792\\
\\
HER-ALL & HER & \multicolumn{1}{r}{1 - 706} & 0 & 2401\\
LER-ALL & LER & 0 & \multicolumn{1}{r}{1 - 999} & 7936\\
LER-VACS & LER & 0 & \multicolumn{1}{r}{1 - 875} & 1238\\
\\
HER-REF & HER &  \multicolumn{1}{r}{3 - 149} & 0 & 62\\
HER-PS & HER & \multicolumn{1}{r}{286 - 449} & 0 & 190\\
HER-VS1 & HER & \multicolumn{1}{r}{216 - 300} & 0 & 265\\
HER-VS2 & HER & \multicolumn{1}{r}{301 - 400} & 0 & 162\\
\\
LER-REF & LER & 0 & \multicolumn{1}{r}{3 - 398} & 220\\
LER-PS & LER & 0 & \multicolumn{1}{r}{185 - 486} & 617\\
LER-VS & LER & 0 & \multicolumn{1}{r}{272 - 350} & 97\\
LER-SA & LER & 0 & \multicolumn{1}{r}{2 - 185} & 239\\
 \bottomrule
\end{tabular}

\label{tab:setup:datasets}
\end{table}

The naming of the data sets reflects the key features, with the first part of the name referring to the type of injection required in the selection, and the second part indicating a specific property to help to differentiate between the sets. Although longer waveforms have also been recorded, all data sets studied here consist of waveforms with a length of \SI{2.4}{\ms} (\SI{3e6}{samples}), since this is the minimum length of all considered data. 

The beam gate of the injector linac, $BG$, as provided by the SuperKEKB beam and condition monitors, is used as an indicator for the type of injection. Three different cases are considered: events recorded during injections into the HER, denoted by ``HER"; events recorded during injections into the LER, denoted by ``LER"; and events during periods without injections, denoted by ``-" in the table and a NI (short for \textit{non-injection}) in the name of the respective data set. The beam currents, also provided by the beam and condition monitors, are given as the range of values among events in the data set after selection. A cut on the beam currents is applied to select events with only one of the two beams filled, or with both beams combined. A beam current above \SI{1}{\mA} in one ring and below \SI{1}{\mA} in the other is required for single ring data sets, whereas a beam current above \SI{1}{\mA} in both rings is required for joint data sets. Finally, $N_{evts}$ denotes the total number of events in the respective data set after the selection.

NI-ALL, NI-HER, NI-LER and NI-VACS are non-in\-jection data sets selected for the study of the structure of backgrounds of regular bunches . Non-injection events are recorded using the random internal auto-trigger without any time correlation to processes in the accelerator (see Section~\ref{sec:setup:setup}).
NI-ALL includes all events with beams in both rings, and thus covers a wide range of partly very different accelerator conditions. The equivalents for events with beams in a single ring are NI-HER and NI-LER.

The term \textit{vacuum scrubbing} describes the outgassing of impurities in the vacuum system of a particle accelerator which is accelerated by the circulating beams via electron-induced and photon-induced desorption. During vacuum scrubbing very high beam currents and artificially inflated beam sizes are used to promote desorption reactions, which also maximizes the backgrounds of ordinary circulating bunches. NI-VACS includes all events with beams in both rings taken (only) during periods without injections and vacuum scrubbing in both rings. As such, it is a subset of NI-ALL.

HER-ALL and LER-ALL are collections of all injection-triggered events (see Section~\ref{sec:setup:setup}) with beams only in the respective ring. LER-VACS is a subset of LER-ALL consisting of events taken during periods of vacuum scrubbing. 

During regular physics operation of SuperKEKB, it is foreseen to simultaneously inject new particles into two bunches separated by 49 RF buckets (\SI{96.3}{\ns}). During Phase~1, double bunch injection was only employed for the LER, and there only in specific cases. One of these cases is vacuum scrubbing for which it was the default injection scheme. Thus, LER-VACS exclusively and LER-ALL partially consist of events with double bunch injections. All other injection data sets are solely based on events with single bunch injections. The HER-ALL, LER-ALL and LER-VACS data sets consist of a large number of events but are taken over a wide range of accelerator conditions.

During Phase~1, SuperKEKB and BEAST~II performed a dedicated injection background study to measure background levels and time structures under controlled conditions, and to assess the effect of variations in the injection parameters. The study was conducted separately for HER and LER injections, and has resulted in four data sets for each beam. First, a reference run with nominal injection parameters was taken as a baseline for both rings; the data sets for the reference injection runs are labeled HER-REF and LER-REF. For each of the other given data sets, one injection parameter was changed from nominal settings to the value yielding the minimal injection efficiency~\footnote{Injections with the highest particle loss rate still consistent with radiation limits throughout the accelerator tunnel.} in order to study its effect on the time structure and on the background levels. For HER-PS and LER-PS, the phase with which particles are injected was changed from \SI{258}{\degree} to \SI{305}{\degree} and from \SI{1}{\degree} to \SI{31}{\degree}, respectively. For HER-VS1, HER-VS2 and LER-VS, we changed the vertical incidence angle from \SI{-0.385}{\milli\radian} to \SI{-0.465}{\milli\radian} and \SI{-0.435}{\milli\radian}, and from \SI{0.123}{\milli\radian} to \SI{0.043}{\milli\radian}, respectively. And for LER-SA, we changed the septum angle from \SI{5.51}{\milli\radian} to \SI{5.39}{\milli\radian}. All of the data sets from the injection study consist of relatively few events but are each taken under very specific and homogeneous accelerator conditions.


\section{Time structure of injection backgrounds}
\label{sec:time_structure}
In this section, the general time structure of injection backgrounds observed with the CLAWS detector is examined. As discussed in Section~\ref{sec:backgrounds}, bunches which recently received new particles due to an injection exhibit drastically elevated particle loss rates. It is thus expected that higher background levels will be observed during transits of these bunches, which recur with every turn until the new particles have fully merged into the original bunch. Since the positron damping ring, which is critical for low-emittance injections of positrons into the main ring, was not yet operational during Phase~1 data taking, injection backgrounds in the LER are expected to be higher than those in the HER. The number of particles detected from non-injection backgrounds is generally low, on the order of $3\times10^4$ particles per bunch, corresponding to less than 1 particle per turn.


\subsection{HER time structure}
\label{sec:time_structure:her}

Due to significant event-by-event fluctuations, we study the time structure of backgrounds by averaging the waveforms of all events in a given data set.

\begin{figure*} 
\centering
\includegraphics[width=\textwidth]{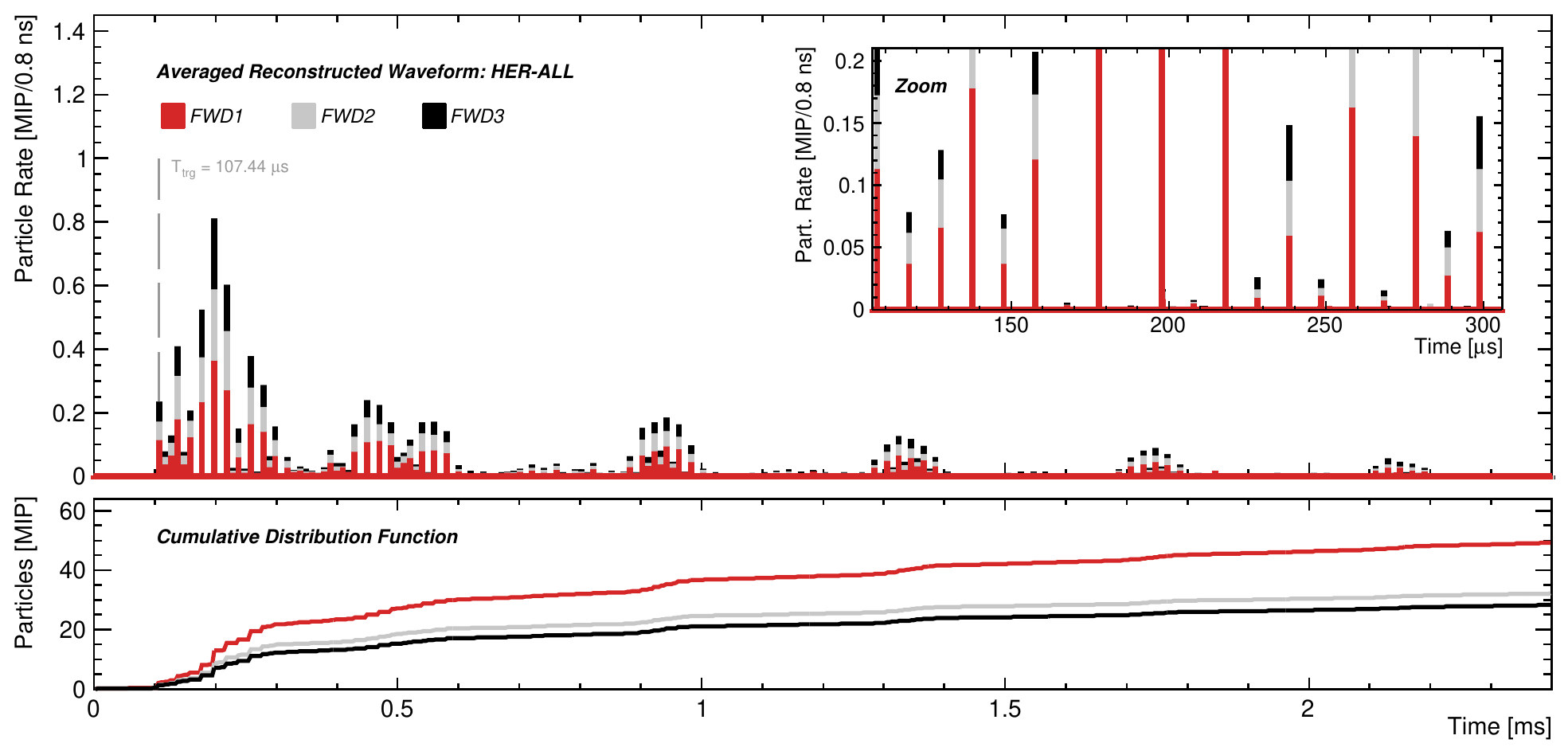}
\caption{Averaged background time distribution for the HER-ALL data set for all three channels stacked on top of each other (top) and their cumulatives separately for each channel (bottom).}
\label{fig:time_structure:her}
\end{figure*}

Figure~\ref{fig:time_structure:her} shows such an averaged waveform for the HER-ALL set for all three channels stacked on top of each other (\textit{top}) and their cumulatives separately for each channel (\textit{bottom}).

As discussed in Section~\ref{sec:setup:setup}, data taking during injections is triggered by an external trigger signal which is sent in advance of the injection. The first injection backgrounds are hence shifted with respect to the start of the waveform by a trigger delay, $T_{trg}$. The exact value of $T_{trg}$ depends on the injection settings used by the accelerator and thus varies slightly in practice.
For the HER, three different values of approximately \SI{107.44}{\micro\second} are observed for the delay which differ on the order of \SI{10}{\nano\second}.
\footnote{We determine the delay by calculating it from the revolution time, $T_{rev}$, and the time in turn, $t_{turn}$, associated with the injection bunch (see Section~\ref{sec:hit_energy}): $T_{trg}=n\times T_{rev} + t_{turn}$, where $n$ is obtained from the averaged waveform. Uncertainties are given by the uncertainties on
$t_{turn}$ since $T_{rev}$ and $n$ are assumed to be exactly known}

After the delay, we find several large peaks of significantly elevated background rates. These peaks are caused by the repeated transit of a single bunch which received new particles in a top-up injection. Thus, the distance between the peaks corresponds to multiples of the beam revolution period ($T_{rev} = \SI{10.0614}{\micro\second}$); this is further highlighted in the inset in the upper right corner. Injection backgrounds are generally declining over time until they return to the levels of regular beam backgrounds and are mostly confined to the first \SI{500}{\us}. The most striking aspect of the observed distribution is that they are not decreasing monotonically with every turn. Instead, the time structure is largely determined by the propagation of the newly injected particles along the beam lines. In that way, the decay behavior is connected to properties of the accelerator and does not have an obvious analytical form.

The background levels observed during transits of the injection bunch are significantly higher than the signals caused by Touschek and beam-gas scattering, which are distributed uniformly over the whole waveform.
In the cumulatives in Figure~\ref{fig:time_structure:her} (\textit{bottom}), larger energy deposits appear as steps, whereas regular backgrounds are reflected by steady increases with a moderate slope.
Injections primarily cause large instantaneous backgrounds which are not necessarily a major contribution to the overall background rates.
As expected, FWD1 shows the largest overall background levels. Distributions for FWD2 and FWD3, on the other hand, are smaller and almost identical.

Other HER data sets generally show a similar behavior. For data sets with few events (see HER-REF, HER-PS, HER-VS1 and HER-VS2 in Table~\ref{tab:setup:datasets}), we observe that the time structure is less pronounced and subject to large turn-by-turn fluctuations. This is because peaks in the time structure are driven by single events with transits with large backgrounds. Absolute background levels and the decay behavior will be discussed more quantitatively in Sections~\ref{sec:hit_energy} and~\ref{sec:decay_behavior}.


\subsection{Time patterns}
\label{sec:time_structure:time_patterns}

In general, we find that three distinct timing patterns occur in the sequence of injection backgrounds: signals occurring every other turn, resulting in a period of 2 turns ($\sim$\SI{20}{\us}), from here on referred to as \textit{betatron pattern}; groups of signals recurring with a period of 9 to 11 turns ($\sim$\SI{100}{\us}), from here on referred to as \textit{long betatron pattern}; and groups of signals recurring with a period of 40 (HER) or 50 (LER) turns, from here on referred to as \textit{synchrotron pattern}. These timing patterns are observed across all injection data sets, with an intensity that varies from data set to data set. The observation of these patterns shows that the time structure of injection backgrounds is strongly affected by betatron and synchrotron oscillations performed by the injected particles.

Time patterns of background signals due to betatron oscillations occur since the trajectories of real beam particles differ from the nominal orbit defined by the lattice of the accelerator. Deviations transverse to the primary direction of motion result in stable horizontal and vertical betatron oscillations around the nominal orbit. In addition, the use of the betatron injection scheme by SuperKEKB, in which new particles are deliberately injected with a horizontal offset (see Section~\ref{sec:backgrounds}), increases these oscillations for newly injected particles.
At the maxima of the oscillations, the particles have an increased probability of colliding with machine elements.
If maxima in the amplitude occur in the proximity of the IP such that scattered or secondary particles can reach the sensors, the oscillations result in elevated background rates.
Since the CLAWS sensors are mounted in the horizontal plane, they are particularly sensitive to backgrounds generated by the oscillations of newly injected particles.
\begin{figure}
\centering
 \includegraphics[width =0.49\textwidth]{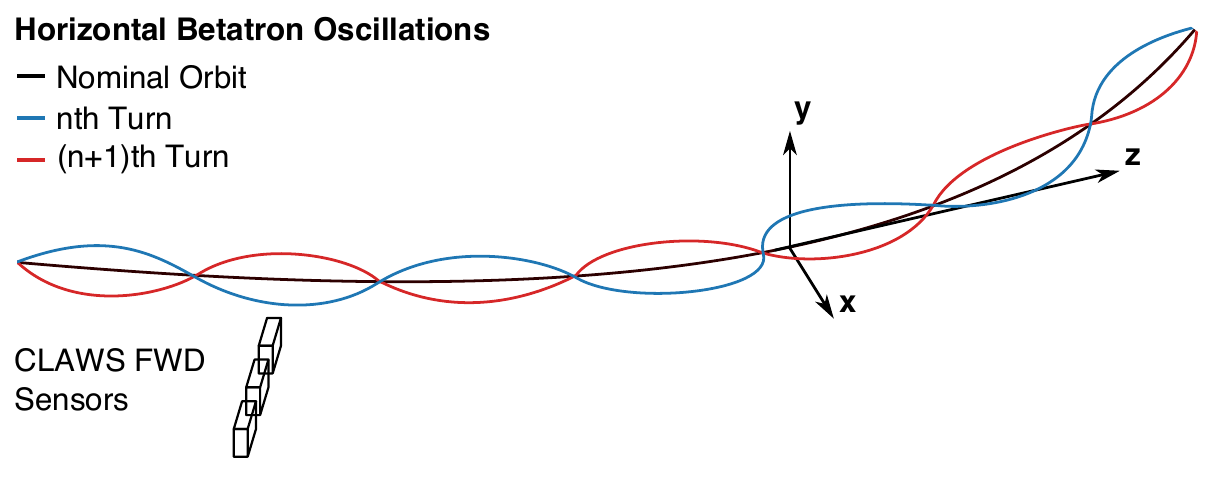}
 \caption{Illustration of the horizontal betatron oscillations of the beam particles relative to the position of the CLAWS FWD sensors.}
 \label{fig:time_structure:oscillations}
 \end{figure}

Figure~\ref{fig:time_structure:oscillations} illustrates the position of the beam particles which are undergoing such oscillations with respect to the CLAWS FWD sensors.

A tracking simulation of beam orbits in the HER, which was performed using the same machine parameters as applied during the reference injections (see HER-REF in Table~\ref{tab:setup:datasets}), predicts horizontal and vertical betatron tunes of 45.53 and 43.57 oscillations per turn, respectively~\cite{Ohnishi:2016yyh}.
Depending on the conditions of the machine, the values of the tunes slightly deviate from the default values.

Since we observe the injection backgrounds from a fixed point and only once per turn, the integer part of the tune is of no relevance for the following discussion and will be disregarded. The fact that the betatron tune is approximately half-integer, on the other hand, can explain the observed \SI{20}{\us} betatron pattern.
If in the $n$th turn the maximum is located at a position where the generated backgrounds reach the sensors it will be shifted strongly away from this position in the $(n+1)$th transit, and vice versa. A good example of this betatron pattern in Figure~\ref{fig:time_structure:her} are the signals of the sixth ($\sim$\SI{157}{\us}), seventh ($\sim$\SI{167}{\us}) and eighth ($\sim$\SI{177}{\us}) transit. Here, two large signals separated by \SI{20}{\us} are observed, which are interleaved by a transit without significant backgrounds.

On closer inspection, the tune differs slightly from the half-integer value by \num{0.03} oscillations per turn. The long betatron pattern is attributed to this deviation from an half-integer value, which results in a slower modulation of the betatron pattern discussed above. In Figure~\ref{fig:time_structure:her}, this is visible in the structure at half a millisecond with maxima at around \SI{450}{\us} and \SI{550}{\us} which are \SI{100}{\us}, or ten turns, apart.

In addition to betatron oscillations, beam particles also undergo synchrotron oscillations caused by deviations in the energy of the beam particles. The difference in energy leads to an altered beam orbit period due to the deflection in the dipole magnets. The particles which deviate in energy therefore also undergo transverse oscillations, which affects the background rates observed by the sensors.

Compared to betatron oscillations, synchrotron oscillations unfold on somewhat longer time scales with periods of several tens of turns. For nominal accelerator parameters, beam orbit simulations for the HER for Phase 1 predict a period of 40.6 turns (\SI{409}{\us}) which in practice again varies by a few turns depending on concrete parameters~\cite{Ohnishi:2016yyh, Lewis:2018ayu}. This is reflected in the time distribution shown in Figure~\ref{fig:time_structure:her}, where significant and extended peaks of backgrounds are separated by around \SI{400}{\us}. The periodic patterns in the data are further explored in an autocorrelation analysis discussed in Section \ref{sec:timing}.

\subsection{LER time structure}
\label{sec:time_structure:ler}

 \begin{figure*}
  \centering
   \includegraphics[width =0.995\textwidth]{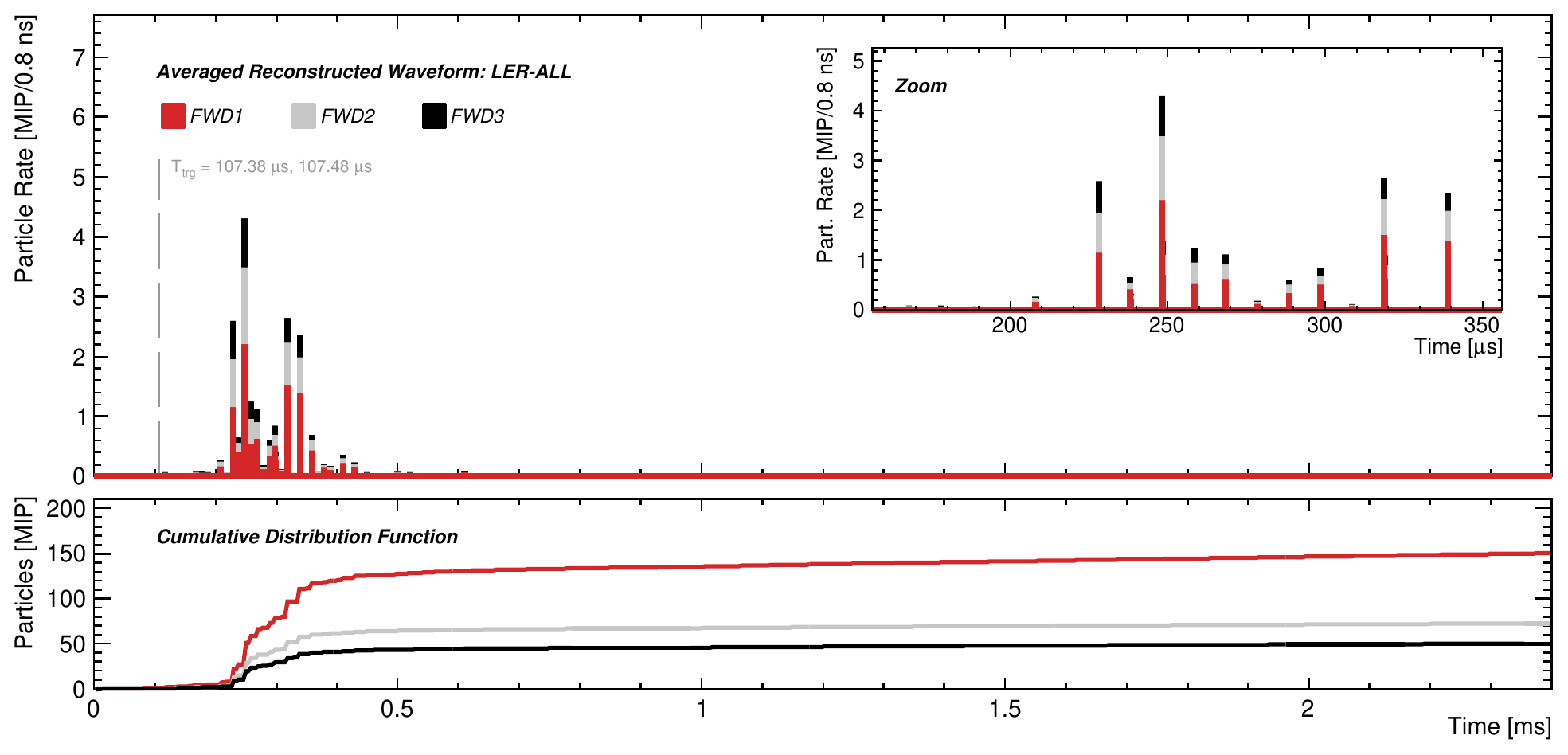}
  \caption{Averaged background time distribution for the LER-ALL data set for all three channels stacked on top of each other (top) and their cumulatives separately for each channel (bottom).}
  \label{fig:time_structure:ler}
  \end{figure*}

Figure~\ref{fig:time_structure:ler} shows the averaged waveform for the LER-ALL data set for all three channels stacked on top of each other (\textit{top}) and their cumulatives separately for each channel (\textit{bottom}). Similar to HER-ALL, several large peaks caused by the repeated transit of the injection bunch are observed. The distances between these signals correspond again to multiples of the $T_{rev}$, particularly evident in the inset in the upper right corner.

The time structure for LER-ALL differs from the one for HER-ALL in a number of respects. The signals are considerably larger in amplitude, and much more concentrated in time.
Small but clear signals (not well visible in Figure ~\ref{fig:time_structure:ler} due to the chosen vertical scale and the line width of the plot) show that the $T_{trg}$ for LER injections is approximately the same as for the HER of \SI{107.4}{\us}. The first significant signals, however, only appear $\sim$\SI{180}{\us} after the start of the recording, i.e. seven full turns after the first transit when taking the time between the injection trigger and the first transit into account.
Instead of the smoother evolution of the background rates observed in HER-ALL, the main particle losses occur in four distinct, non-adjacent transits. No significant energy deposits are observed beyond \SI{500}{\us} after the injection. The most likely explanation for this behavior is the absence of the damping ring for positrons, which leads to higher loss rates for the newly injected particles. This also influences the observable time patterns originating from oscillations of the beams.

For nominal accelerator parameters (see LER-REF in Table~\ref{tab:setup:datasets}), beam orbit simulations for the LER for Phase 1 predict horizontal and vertical betatron tunes of 44.53 and 46.57 oscillation per turn, as well as a synchrotron oscillation period of 52.3 turns (\SI{526}{\us})~\cite{Ohnishi:2016yyh, Lewis:2018ayu}. As for the HER, these values vary with the exact beam parameters.
Note that the non-integer part of the betatron tunes is identical to the prediction for the HER. We observe multiple instances of the betatron pattern, most clearly seen in the inset where the distances between signals is mostly \SI{20}{\us}. The distance between the two largest peaks corresponds to \SI{90}{\us} which is a consequence of the long betatron pattern, which occurs at a slightly changed frequency compared to the HER. The same timing features are also observed in other data sets. No indications for synchrotron oscillations are observed, since the overall duration of the elevated background levels is not sufficiently long for these longer oscillations to emerge.

The absence of the damping ring for positrons leads to larger injection backgrounds in the LER than in the HER, with observed instantaneous background levels as much as five times higher. The radial dependence, in contrast, is comparable to the one observed for HER injections: the signals in FWD1 are larger than in FWD2 which are larger than in FWD3. The background from regular circulating bunches is similar to the one observed in HER injection and non-injection events; it is uniformly distributed and the overall height is negligible compared to the signals from the injected bunch. The overall background levels in the cumulatives are dominated by injection backgrounds, which appear as steps of large energy deposits. The total number of observed particles is an order of magnitude larger than in the HER and driven almost entirely by injection backgrounds.

The averaged waveforms for the other LER injection data sets generally show a background time structure with similar features and overall background levels. For sets with a small number of events (LER-REF, LER-PS, LER-VS and LER-SA in Table~\ref{tab:setup:datasets}), large fluctuations in the structure are observed. LER-VACS was recorded during vacuum scrubbing and using double bunch injections. Here, we find that the majority of backgrounds comes from Touschek and Beam-gas interactions of regular bunches and that there is no noticeable difference between the behavior of the two injection bunches. For double bunch injections into the LER, the separation of the two injection bunches has been measured by CLAWS to be \SI{96.8}{ns} \cite{Lewis:2018ayu}, with a trigger delay relative to the first bunch measured at \SI{107.38}{\micro\second}.

\section{Hit energy spectra}
\label{sec:hit_energy}

As a next step, the hit energies in the detectors are studied. 
First the analysis methodology is introduced and then the hit energy distributions for backgrounds recorded during injection and non-injection periods are discussed.

\subsection{Methodology}
\label{sec:hit_energy:methodology}

Instead of averaging the waveforms as done for the study of the time structure discussed in the previous section, here an event-by-event analysis of each individual waveform of a data set is performed to identify signals from the interaction of a single particle, or from the simultaneous interaction of multiple particles that originate from the same particle bunch in the accelerator. Such a signal is referred to as a \textit{hit}. Since the emission and detection of the scintillation photons is not instantaneous, the capturing of the full energy to be associated to a hit requires to reconstruct clusters of detected photons in time. Clusters are formed by identifying photon deposits within an integration window of 76.8 ns from the first observed photon. The amplitude of the hit is given by the integral over the full duration of the cluster with additional corrections for noise and pedestal uncertainties, while the time is determined from a constant fraction of 0.1 of the total integral. The parameters were obtained from an optimisation taking stability, precision and probability for the identification of fake hits into account \cite{Gabriel2019} In the following, the distribution of the reconstructed energies of hits is referred to as \textit{hit energy distribution}.

To cleanly separate background contributions of the injection bunch from the contributions of other processes, an additional time-based variable referred to as the time in turn, $t_{turn}$, is introduced. It represents the time within one turn of the accelerator, relative to an arbitrary starting time given by the external trigger,
\begin{equation}
t_{turn} = t \bmod T_{rev},
\end{equation}
where $t$ is the absolute time after the start of the waveform and $T_{rev}$ is the revolution time of the accelerator. Over multiple turns, all transits of a particular particle bunch thus correspond to the same $t_{turn}$. 

To enable quantitative comparisons, the distributions discussed in the following are typically normalized to the number of events, to the length of the waveforms in time and to the area of the sensors.
Since the detectors are orthogonal to the beam direction, the integral of the distributions represents the particle flux through the respective CLAWS sensor.


\subsection{Time in turn and time resolved hit energy spectra}
\label{sec:hit_energy:time_resolved_spectra}

\begin{figure}
\centering
\includegraphics[width=\columnwidth]{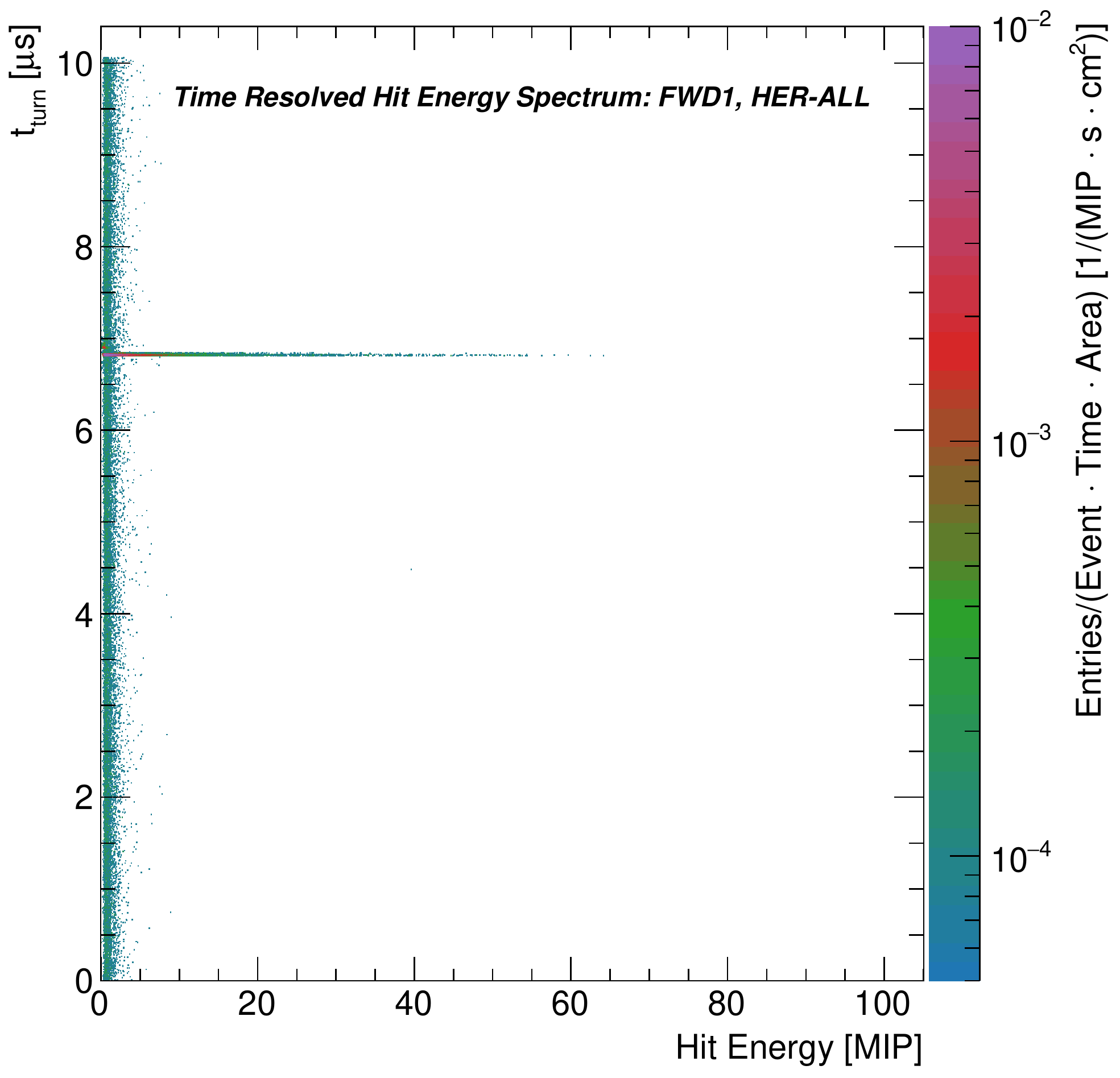}
\caption{The distribution of the time in turn versus the hit energy for the 
HER-ALL data set for channel FWD1. The distribution is normalized to the number of events, the length of the waveforms and the area of the sensors.}
\label{fig:hit_energy:time_resolved_spectra}
\end{figure}

Figure~\ref{fig:hit_energy:time_resolved_spectra} shows the distribution of hit energies for the time in turn versus the hit energy for the HER-ALL data set. The $t_{turn}$ of the injection bunch clearly stands out in this distribution as a narrow peak extending to high hit energies.
The rate of hits with small reconstructed energies below \SI{2}{MIP} is significantly elevated compared to the surrounding area which shows regular beam backgrounds.
Hit energies above \SI{2}{MIP} occur almost exclusively with a $t_{turn}$ corresponding to the injection bunch. This shows that hits caused by multiple simultaneous particles in the CLAWS sensors are essentially confined to the injection bunch, while signals for the other bunches are characterized by the sporadic detection of single charged particles and other low-amplitude signals. 

\begin{figure*}
\centering
\includegraphics[width=0.49\textwidth]{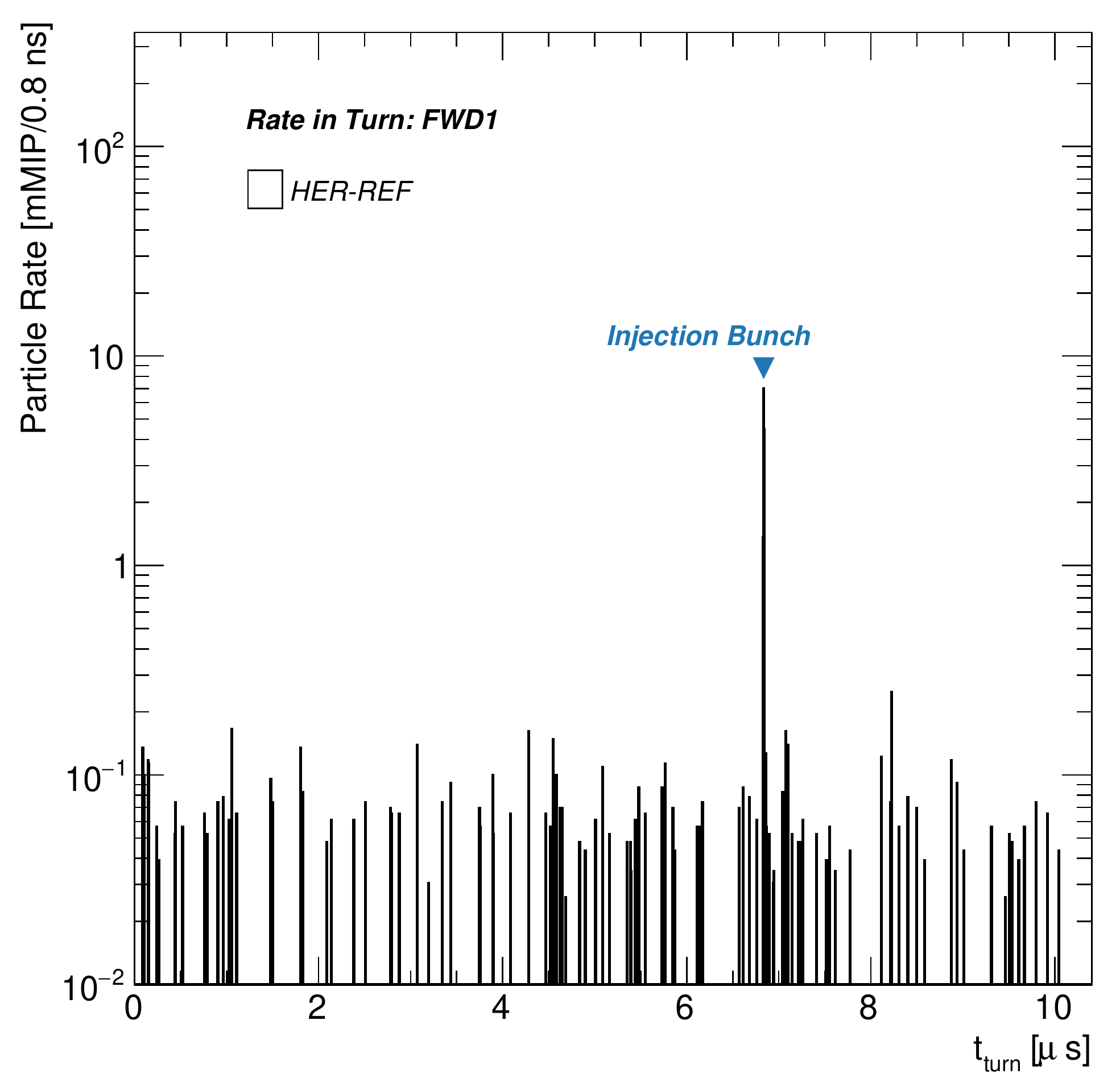}
\hfill
\includegraphics[width=0.49\textwidth]{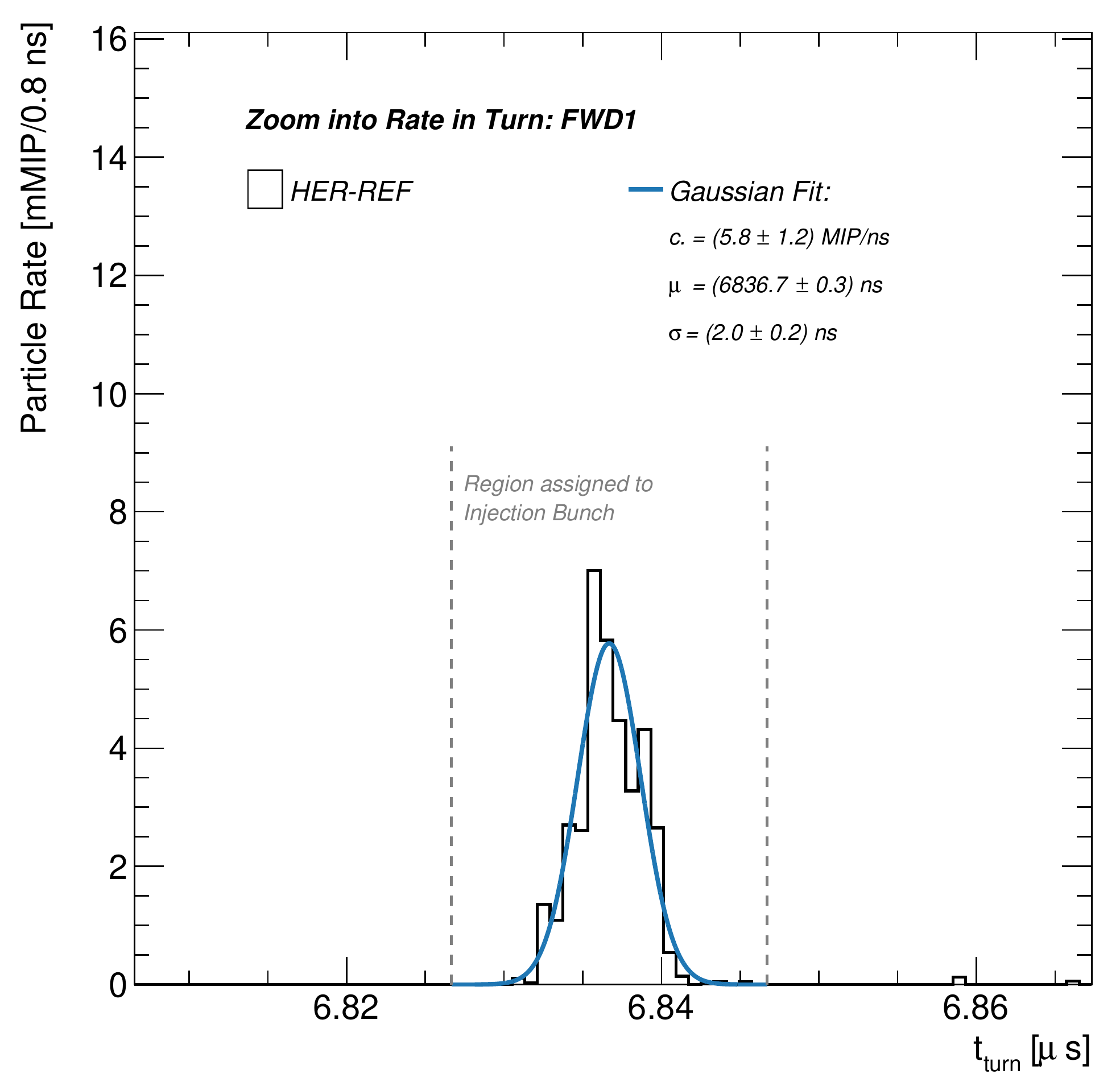}
\caption{Particle rate mapped onto the time in turn for the HER-REF dataset. The left panel shows the rate over the full turn, while the right panel shows a \SI{60}{ns} window around the location of the injection bunch. The time window used to assign energy deposits to the injection bunch, given by $\pm 5 \sigma$ around the maximum of the fitted Gaussian, is also indicated.}
\label{fig:RateInTurn_IB}
\end{figure*}

For a more detailed study of the energy distribution of injection backgrounds the averaged particle rate as a function of the $t_{turn}$ is used instead of the distribution shown in Figure \ref{fig:hit_energy:time_resolved_spectra}. For each data set, the $t_{turn}$ corresponding to the injection bunch is determined by fitting a Gaussian distribution to the peak caused by the elevated particle rates due to the injection background.
This is illustrated in Figure \ref{fig:RateInTurn_IB}, which shows the detected particle rate mapped onto $t_{turn}$ for the HER-REF dataset. The right panel shows the region around the injection bunch, with the Gaussian fit used to extract the position and width.
In the subsequent analyses, all hits which lie within 5\,$\sigma$ of the fitted mean are classified as injection backgrounds.
In this concrete case, the fitted $\sigma$ of \SI{2}{ns} results in a time window of $\pm \SI{10}{ns}$ that is used to assign particles to the injection bunch. For data sets that combine runs from the full time period studied here, such as HER-ALL, two small changes of the injection trigger offset occurred due to changes made by the machine group. This results in three different times for the $t_{turn}$ of the injection bunch, with variations of approximately $\pm \SI{10}{ns}$. For these data sets, each of the three peaks is fitted separately, and the combined time window for all three is used to identify the injection bunch. In these cases, the time window increases from a width of \SI{20}{ns} to approximately \SI{40}{ns}. This does not have a noticeable influence of the results.

\subsection{Hit energy spectra comparison}
\label{sec:hit_energy:spectra}
\begin{figure} 
\centering
\includegraphics[width=\columnwidth]{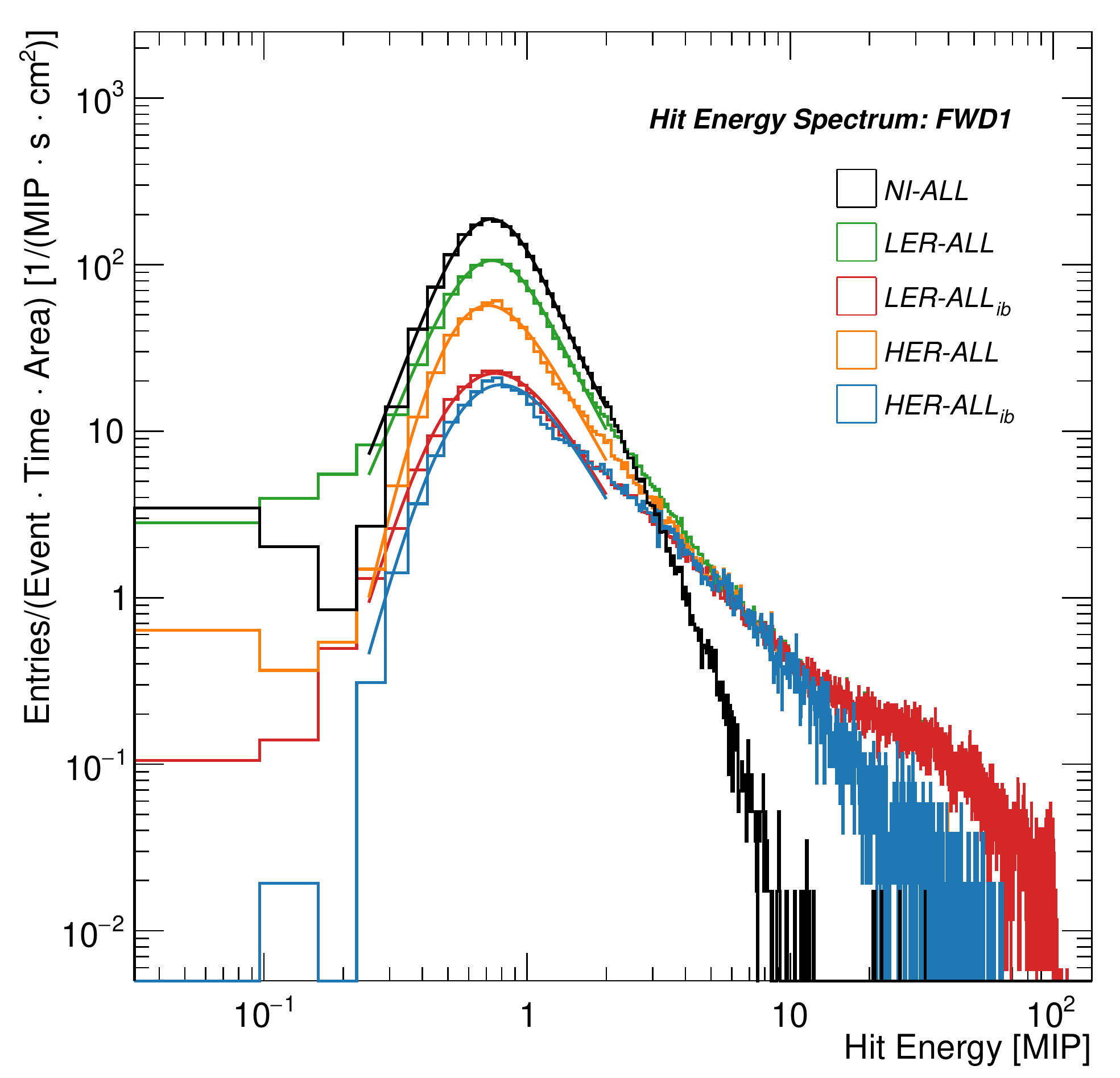}
\caption{Distributions of hit energies for the given data sets for channel FWD1. The distributions are normalized to the length of the waveforms, the area of the sensors and the number of events. For the distributions HER-ALL$_{ib}$ and LER-ALL$_{ib}$ a $t_{turn}$-based timing cut is applied to select only hits caused by the injection bunch.}
\label{fig:hit_energy:spectra}
\end{figure}

Figure~\ref{fig:hit_energy:spectra} compares the hit energy spectra for the NI-ALL, HER-ALL and LER-ALL data sets. It also shows the corresponding distributions of hits caused only by the injection bunches selected by a cut on the $t_{turn}$, as described in the previous section.
We refer to these subsets as HER-ALL$_{\mathrm{ib}}$ and LER-ALL$_{\mathrm{ib}}$.

The probability of observing Touschek and beam-gas backgrounds during the transit of a regular bunch is estimated by calculating the number of hits per bunch for the non-injection data set NI-ALL.\footnote{The number of hits per bunch is calculated by multiplying the particle flux, i.e. the integral of the distribution, by the area of the sensors and the time difference between two consecutive bunches ($3 \times T_{bucket} = \SI{5,895}{\nano\second}$~\cite{Lewis:2018ayu}.)}
Here it is found that on average only around one in ten thousand bunches generates an observable signal, which leads to the conclusion that non-injection backgrounds are caused almost exclusively by single particles.
The distribution hence reflects the functional form expected for a single charged particle, which is also observed in calibration measurements using cosmic muons. The characteristic peak of this distribution can therefore be described by a convolution of a Landau and a Gaussian distribution. 

For hit energies up to \SI{2}{MIP}, all distributions show a similar shape. Since the NI-ALL, HER-ALL and LER-ALL data sets have been taken under vastly different accelerator conditions, the total number of hits between these sets cannot be compared. The difference in the number of hits between HER-ALL and HER-ALL$_{ib}$ or LER-ALL and LER-ALL$_{ib}$, on the other hand, shows that the majority of the hits in this energy range stem from Touschek and Beam-gas interactions. Nevertheless, the injection bunch is only one (or two in the case of double bunch injection events) of 1576 bunches, but contributes around 10\% of the hits.

The similarity in the shape of the distributions suggests that injection backgrounds in this range are also caused by single charged particles and can be described by convolutions of a Landau and a Gaussian distribution. Since the sensors have been calibrated~\footnote{Performed after the run time of Phase~1.} using cosmic muons, the most probable value (MPV) of such a fit should correspond to an energy of \SI{1}{MIP}. For a given channel, the values of the MPVs are consistent across all data sets shown in Figure~\ref{fig:hit_energy:spectra} with mean values and corresponding sample standard deviations of \SI{0.75 \pm 0.03}{MIP} (FWD1), \SI{0.93 \pm 0.03}{MIP} (FWD2) and \SI{1.10 \pm 0.02}{MIP}. The relatively larger deviation from the calibration value for FWD1 is attributed to the fact that the sensor for this channel was most exposed to radiation, and thus has increased uncertainties in the calibration with cosmic muons after the Phase~1 measurements as discussed in Section \ref{sec:setup:setup}. In addition, the reconstructed hit energy for small signals in the MIP range is affected by uncertainties in the pedestal determination as part of the photon signal reconstruction due to the high dark count rate, which increases the probability to miss true photon signals in the tails of the time distribution when the signal merges back into the baseline. The FWD1 sensor was replaced during the Phase~1 run time, which is an additional source of systematic uncertainties on the average detector response.

At energies of approximately \SI{2}{MIP}, the distributions for the injection data sets and NI-ALL are beginning to diverge. For amplitudes beyond this point, it is more likely that hits are caused by injection backgrounds than by beam-gas or Touschek scattering. Above \SI{5}{MIP}, the distributions for the injection bunch alone account for more than 90\% of the observed signals. The distribution for NI-ALL shows a power law behavior with a steady decrease to higher hit energies, reaching negligible rates beyond \SI{12}{MIP}. The highest hit energies observed are \SI{33}{MIP}. The distribution for HER-ALL is following a similar power law, but its decline is less steep in particular once particles from the injection bunch dominate beyond \SI{5}{MIP}, and it reaches considerably higher energies of up to \SI{64}{MIP}. LER-ALL shows a small ankle at around \SI{10}{MIP} for which the decline slows down slightly.
The spectrum is cutting off at around \SI{113}{MIP}, which is due to the limited dynamic range of the CLAWS system. Here two effects come into play, the limit of the analog range of the amplifier chain of the system, and the saturation of the SiPM due to the finite number of pixels of the photon sensor. The former effect truncates signals that exceed an amplitude corresponding to approximately 450 p.e.\ simultaneously. The reconstructed amplitude of such signals then only grows with the length of the signal in time above the cut-off, rather than with its full amplitude. The SiPM saturation itself results in a non-linear behaviour of the detection of the scintillation photons, reducing the sensor response for large signals. Both effects limit the maximum hit energy that can be observed, shifting larger signals to lower reconstructed hit energies. This results in an increased rate observed close to the highest energy, visible as a bump at the end of the spectrum.

While the saturation effects are clearly noticeable in the hit energy spectrum, they only affect a very small fraction of the observed hits. The dynamic range of the CLAWS system approximately matches the largest signal amplitudes observed, with only few events on a steeply falling spectrum entering into the range where saturation effects become relevant. Saturation thus has a negligible impact on more global variables discussed in the next section. This is also true for LER-PS, the subset of LER-ALL with the largest overall background levels, where saturation effects are most pronounced. As a consequence, no saturation corrections are applied in the analysis of Phase~1 data.

The most striking result to emerge from the data is that the distributions of HER-ALL$_{\mathrm{ib}}$ and LER-ALL$_{\mathrm{ib}}$ are merging with HER-ALL and LER-ALL, respectively, at hit energies above 3 -- 4 MIP. This shows that large instantaneous background levels with multiple particles hitting the CLAWS sensors simultaneously are caused exclusively by injection bunches.

Qualitatively, similar distributions are also observed for the channels further away from the beam line, FWD2 and FWD3. The overall number of hits is generally decreasing with the distance from the beams as expected from the radial profile of the sensor positions (FWD1 > FWD2 > FWD3). Across all injection data sets, significantly elevated instantaneous background levels are observed during transits of the injection bunches. The highest injection background levels observed during Phase~1 do not exceed the levels of regular beam backgrounds by more than an order of magnitude. These results, together with other BEAST~II results, have contributed to the assessment that the injection background is manageable, and the overall radiation levels are safe for the installation of Belle~II for the subsequent commissioning phases.


\section{Overall injection background levels and decay behavior}
\label{sec:decay_behavior}

Understanding the impact of injection backgrounds on the operation of the Belle~II detector requires detailed knowledge of the overall particle rates and the duration of the period of elevated backgrounds which follow an injection. In this section, a method for quantifying these deposited energies and the decay behavior of injection backgrounds with respect to accelerator conditions is discussed.


\subsection{Methodology}
\label{sec:decay_behavior:methodology}

\begin{table*}
\centering
\caption{Key quantities summarizing injection background energy levels and decay behavior for all injection data sets for channel FWD1. Values are stated by the median of all the values in the data set; subscripts denote the difference from the median to the \SI{16}{\percent} quantile, whereas superscripts indicate the difference from the median to the \SI{84}{\percent} quantile.}
\begin{tabular}{l c c c c c c c c}
\toprule
Data Set & \phantom{}& {$E$ [MIP]} & \phantom{} & {$E_{ib}$ [MIP]} & \phantom{}& {$E_{ib}$/$E$ [\%]} & {$E_{ib}$($t<$\SI{0.5}{\ms})/$E_{ib}$ [\%]} & {$t(E_{ib}<$\SI{2}{MIP}) [\si{\us}]}\\
\midrule
\addlinespace[6pt] 
HER-ALL && $27_{-19}^{+81}$ && $10_{-10}^{+94}$ && $55_{-53}^{+43}$ & $64_{-64}^{+27}$ & $292_{-292}^{+1469}$\\ \addlinespace[6pt]
LER-ALL && $123_{-82}^{+147}$ && $80_{-78}^{+91}$ && $86_{-77}^{+11}$ & $99_{-11}^{+1}$ & $359_{-201}^{+241}$\\ \addlinespace[6pt]
LER-VACS && $41_{-32}^{+50}$ && $2_{-2}^{+8}$ && $6_{-6}^{+70}$ & $100_{-41}^{+0}$ & $0_{-0}^{+369}$\\ \addlinespace[6pt]
\\
HER-REF && $11_{-9}^{+7}$ && $9_{-8}^{+7}$ && $86_{-25}^{+14}$ & $81_{-33}^{+19}$ & $141_{-141}^{+322}$\\ \addlinespace[6pt]
HER-PS && $48_{-16}^{+18}$ && $32_{-16}^{+13}$ && $65_{-19}^{+8}$ & $76_{-26}^{+14}$ & $775_{-483}^{+493}$\\ \addlinespace[6pt]
HER-VS1 && $15_{-6}^{+11}$ && $3_{-3}^{+11}$ && $22_{-22}^{+36}$ & $78_{-58}^{+22}$ & $141_{-141}^{+0}$\\ \addlinespace[6pt]
HER-VS2 && $26_{-10}^{+12}$ && $7_{-6}^{+11}$ && $30_{-23}^{+24}$ & $78_{-78}^{+19}$ & $141_{-141}^{+0}$\\ \addlinespace[6pt]
\\
LER-REF && $144_{-21}^{+15}$ && $138_{-20}^{+12}$ && $96_{-4}^{+2}$ & $100_{-1}^{+0}$ & $379_{-20}^{+50}$\\ \addlinespace[6pt]
LER-PS && $262_{-104}^{+23}$ && $242_{-92}^{+21}$ && $95_{-3}^{+2}$ & $99_{-1}^{+1}$ & $590_{-141}^{+91}$\\ \addlinespace[6pt]
LER-VS && $176_{-15}^{+20}$ && $162_{-14}^{+19}$ && $94_{-3}^{+2}$ & $99_{-1}^{+1}$ & $409_{-50}^{+201}$\\ \addlinespace[6pt]
LER-SA && $181_{-15}^{+13}$ && $176_{-13}^{+14}$ && $99_{-1}^{+1}$ & $97_{-2}^{+2}$ & $661_{-302}^{+20}$\\ \addlinespace[6pt]
 \bottomrule
\end{tabular}
\label{tab:decay_behavior}
\end{table*}

As mentioned in the previous two sections, three main features of injection backgrounds are: the time evolution does not follow a well defined functional form; the background levels and the time structures fluctuate substantially between individual events of a data set; and the background levels and the time structures also fluctuate substantially for different accelerator conditions represented by the different data sets. For Phase~1 of SuperKEKB commissioning, injection backgrounds can therefore not be described analytically. 

Instead, the behavior of injection backgrounds is characterized based on the following quantities which are sensitive to the total and injection background levels and the time evolution of the injection background:
\begin{itemize}
    \item the total energy deposited in a single event including regular beam backgrounds, $E$, given by the sum of all hit energies in the event;
    \item the energy deposited only by the injection bunch, $E_{ib}$, given by the sum of the energies of all hits selected by the cut on $t_{turn}$ (see Section~\ref{sec:hit_energy:time_resolved_spectra});
    \item their ratio $E_{ib}$/$E$;
    \item the fraction of the energy deposited by the injection bunch within \SI{500}{\us} after the first transit,
\footnote{This range is chosen because the analysis in Section \ref{sec:time_structure} shows that injection backgrounds are often confined to the first \SI{500}{\us}.}
$E_{ib}(t<$ \SI{500}{\us})/$E_{ib}$;
\item and the time of the last transit in which the $E_{ib}$ per transit is larger than \SI{2}{MIP}, $t(E_{ib} >$ 2 MIP).\footnote{For the analysis of Phase~1 data, \SI{2}{MIP} is chosen as a limit because it is approximately the energy at which the hit energy spectra of injection and non-injection backgrounds begin to diverge (see Section~\ref{sec:hit_energy:spectra}).}
\end{itemize}    
These quantities are determined independently for each event.

No notable differences are observed in the behavior of the first and second injection bunch for data sets which include events recorded during double bunch injections (i.e. LER-ALL and LER-VACS, see Section~\ref{sec:setup:datasets}). Contributions from the second bunch are selected by a dedicated timing cut and subtracted from the quantities in order to make all data sets comparable.

For the HER-ALL and LER-ALL data sets, the large number of events allows to determine meaningful distributions of the quantities. For all other sets, the quantities are stated by the median; the difference between the median and the \SI{16}{\percent} quantile (denoted in subscript); and the difference between the median and the \SI{84}{\percent} quantile (denoted in superscript). The values obtained for the quantities for all injection data sets are summarized in Table~\ref{tab:decay_behavior}. The detailed distributions of the quantities for the HER-ALL and LER-ALL data sets are introduced in the following, prior to a discussion of  the findings for all data sets.

\subsection{Total and relative deposited energies}

\begin{figure}
	\centering
    \includegraphics[width =0.99\columnwidth]{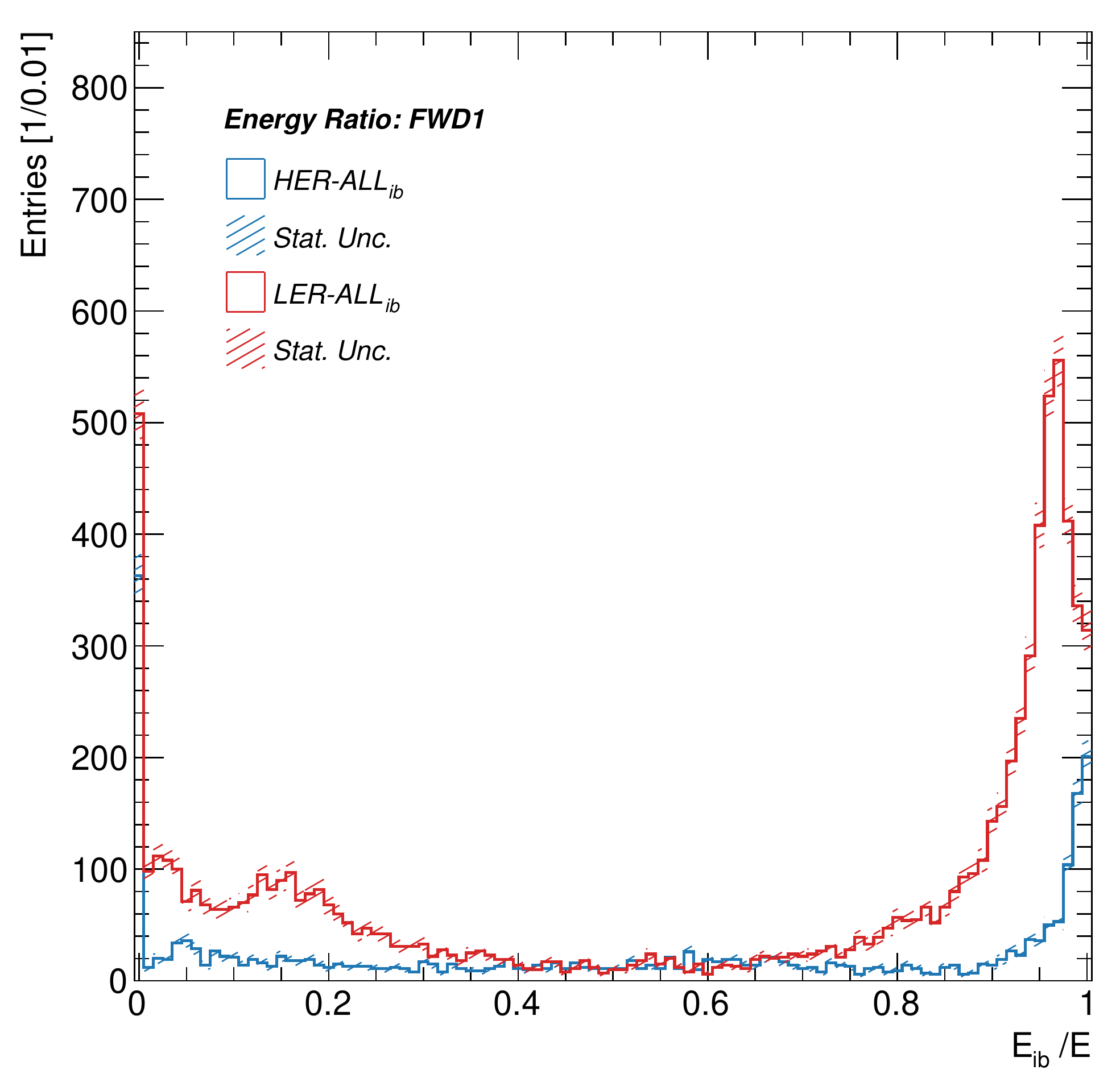}
    \caption{Distributions of the ratios of the energy deposited only by the injection bunch, $E_{ib}$, to the total deposited energy, $E$, for channel FWD1 for HER and LER injection runs. Events without any detected backgrounds are ignored (\SI{0.8}{\percent} and \SI{1.6}{\percent} for HER-ALL and LER-ALL, respectively).}
    \label{fig:decay_behavior:energies}
\end{figure}

As expected, the total deposited energy in LER-All is considerably larger than for HER-ALL.
Comparing $E_{ib}$ with $E$ for both data sets shows that the majority of the observed backgrounds is caused by the injection bunch.
The fraction of the total observed energy originating from the injection bunch, $E_{ib}$/$E$, is examined on an event-by-event basis in Figure~\ref{fig:decay_behavior:energies}.

Events in which no background hits are observed are ignored (\SI{0.8}{\percent} and \SI{1.6}{\percent} for HER-ALL and LER-ALL, respectively). In a small fraction of the events (\SI{15.2}{\percent} and \SI{6.5}{\percent} for HER-ALL and LER-ALL, respectively), the injection bunch did not deposit any energy in the sensors, which leads to large peaks at zero. These events can be explained by injections with low efficiency leading to severe instantaneous particle loss before the bunch reaches the IP, or situations where the particle loss takes place only at locations other than the IP. They could also represent events which are falsely classified as injection events in the event selection (see \textit{BG} in Section~\ref{sec:setup:datasets}).

There is a relatively uniform distribution of values for ratios between 0.2 and 0.8, with a strong rise and subsequent peaks around 0.95 for the LER, and 1 for the HER. These are caused by events with low regular beam backgrounds and/or large injection backgrounds.

\subsection{Injection background decay time}

\begin{figure}[t]
	\centering
    \includegraphics[width =0.99\columnwidth]{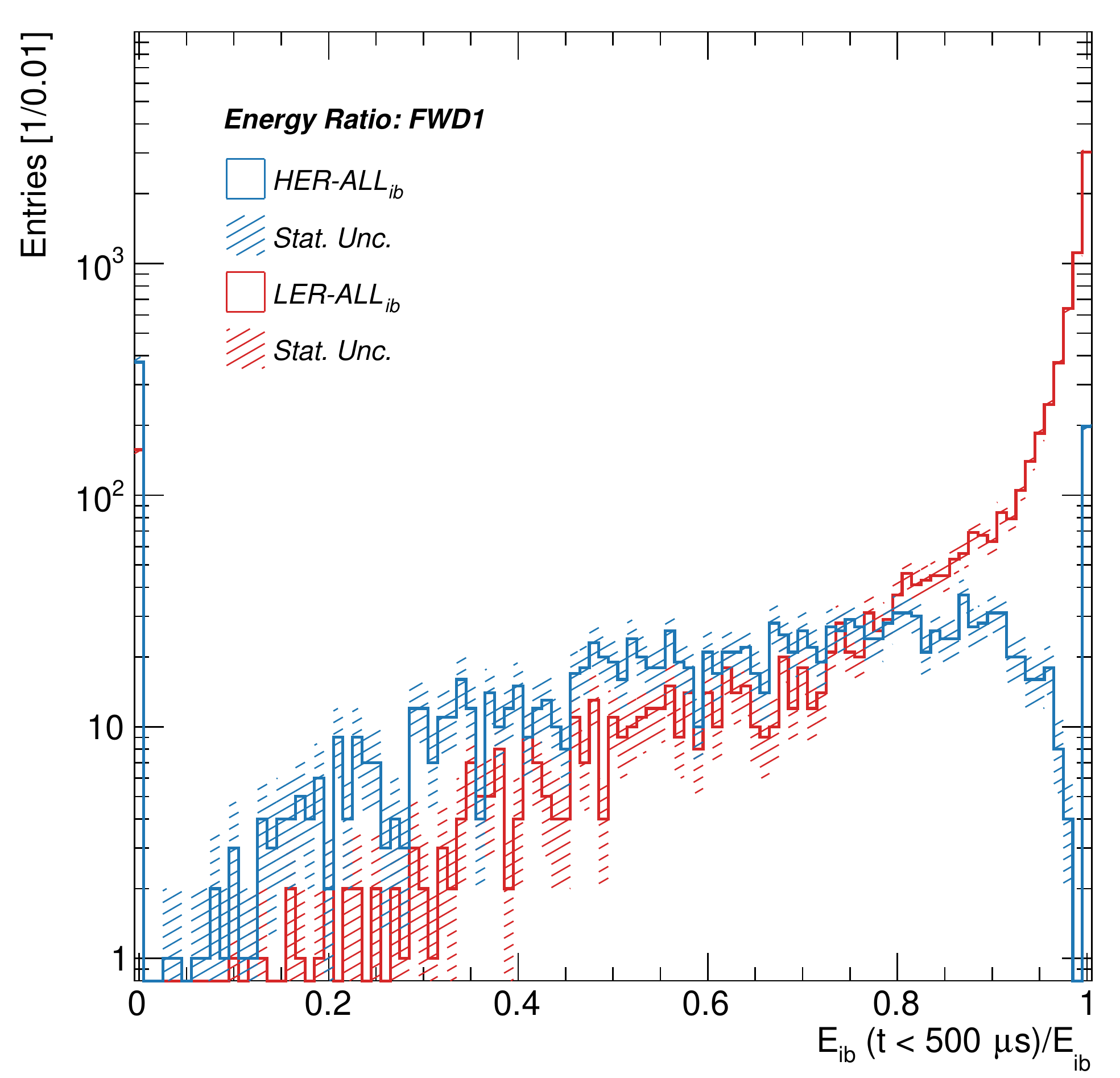}
    \caption{Distributions of the ratios of the energy deposited within the first \SI{500}{\us} after the first transit, $E_{ib}$($t<$\SI{0.5}{\us}), to the full energy deposited by the injection bunch, $E_{ib}$, for channel FWD1 for HER and LER injection runs. Events without any injection backgrounds are disregarded (\SI{15.9}{\percent} and \SI{7.9}{\percent} for HER-ALL and LER-ALL, respectively).}
    \label{fig:decay_behavior:decay_time}
\end{figure}

The time it takes for the background to reach normal levels after an injection is a critical parameter for the operation of the Belle~II detector. The absence of focusing elements and other accelerator elements that are present for physics running significantly changes the timing patterns of the injection background, as discussed in Section \ref{sec:time_structure}. Nevertheless, the evolution of the background levels over many revolutions of the accelerator can provide important insights. To make this accessible despite the large turn-to-turn fluctuations in the background level, the fraction of the energy which is deposited within the first \SI{500}{\us} to the total energy deposited by the injection bunch, $E_{ib}(t<$ \SI{500}{\us}), is studied. 

Figure~\ref{fig:decay_behavior:decay_time} shows this variable for the HER-ALL and LER-ALL data sets. Events without any injection backgrounds (\SI{15.9}{\percent} and \SI{7.9}{\percent} for HER-ALL and LER-ALL, respectively), are ignored, but  events in which the injection bunch did not cause any hit within the first \SI{500}{\us} (\SI{18.6}{\percent} and \SI{2.2}{\percent} for HER-ALL and LER-ALL, respectively) are included, which results in the peak at 0. Neglecting the edges, the distribution for the HER-ALL set shows a relatively uniform distribution with a skewness to larger ratios. This clearly shows that injection backgrounds in the HER are only partially confined to the first \SI{500}{\us}. The distribution for LER-ALL, on the other hand, is peaking at one and is declining steeply to smaller ratios. Injection backgrounds in the LER are thus decaying faster and are mostly confined to the fist 50 turns.

\subsection{Time of the last hit}

\begin{figure}[!tb]
  \centering
  \includegraphics[width =0.99\columnwidth]{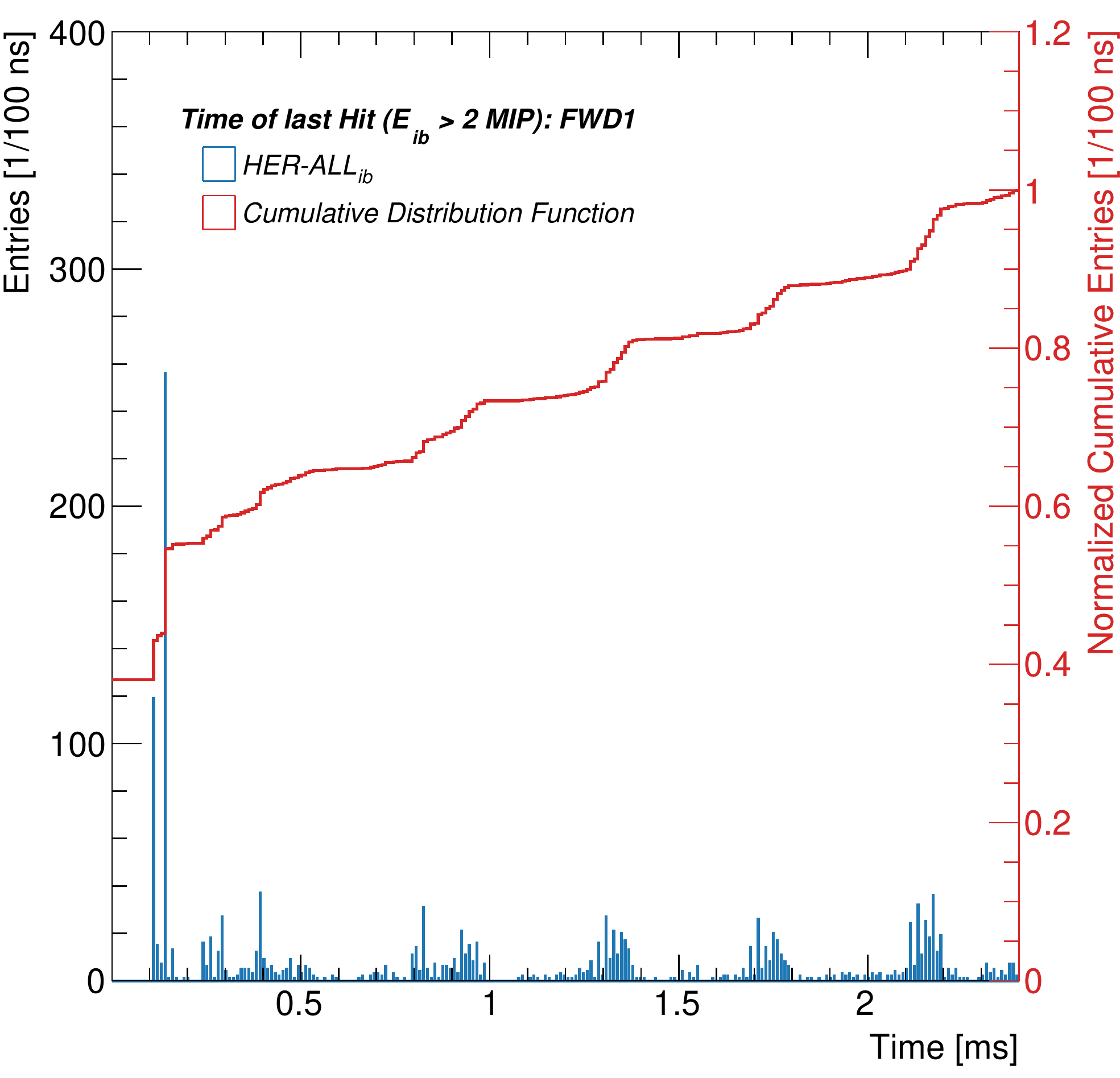}
  \caption{The distribution of the time of the last transit in which the energy per transit was larger than \SI{2}{MIP} and the corresponding cumulative distribution function for channel FWD1 for the HER-ALL data set.
  Distribution of the time of the last transit (not cumulative) does not show events in which the injection bunch did not cause any hit above the threshold (\SI{42.3}{\percent}).}
  \label{fig:injection:injection:decay:tbelow}
\end{figure}

Another parameter that provides information of the temporal extension of elevated background levels following particle injection is the time of the last transit in which the energy deposited by the injection bunch is above a predefined limit. For the analysis of Phase~1 data, \SI{2}{MIP} is chosen as a limit since this is the energy at which the hit energy spectra of injection and non-injection backgrounds begin to diverge (see Section~\ref{sec:hit_energy:spectra}). Thus, $t(E_{ib}<$ \SI{2}{MIP}) indicates the time after which injection backgrounds are comparable to regular beam backgrounds. In principle it is possible that a hit, which is recorded during a transit and exceeds the limit, is caused by regular beam backgrounds and not by injection backgrounds.
The hit energy spectra for non-injection backgrounds, however, suggest that such a hit occurs in less than 1.1\% of the events.

Figure~\ref{fig:injection:injection:decay:tbelow} shows the distribution of the time of the last transit and the corresponding cumulative distribution function for the HER-ALL data set. The observations are generally in agreement with the findings from the fractions of $E_{ib}(t<$ \SI{500}{\us}). In \SI{42.3}{\percent} of the events, the injection bunch did not cause any hit above the threshold (disregarded in the distribution, not in the cumulative). The two largest peaks are located directly at the first and fourth transit of the injected bunch. Apart from that, the distribution is stretching across the full \SI{2.4}{\ms} with periodically repeating patterns attributed to synchrotron oscillations of the beam particles (see Section~\ref{sec:time_structure:time_patterns}).
This indicates that the timescale of injection backgrounds in the HER is fluctuating strongly from injection to injection, and that in a significant fraction of all injections a full decay requires at least \SI{2.4}{\ms}.

For the majority of the events of the LER data sets, the decay is taking place entirely within \SI{700}{\us} after the first transit. In a small number of events, however, injection backgrounds are also long lasting, showing influences of synchrotron oscillations.

\subsection{Overall results and injection studies}

Following the discussion of individual variables for selected high-statistics data sets, the quantities obtained for all injection data sets are compared (summarized in Table~\ref{tab:decay_behavior}). Particular emphasis is placed on the sets which are part of the dedicated injection studies. The values determined for the overall background, $E$, range from \si{11} to \SI{262}{MIP}, with LER data sets showing significantly higher energies than HER sets. The corresponding spread indicates large event-to-event fluctuations across all data sets, which can be more than \SI{100}{\percent} of the median total energy. Comparing the $E_{ib}$ with $E$ suggests that in all data sets except LER-VACS, a major part of the observed backgrounds is caused by the injection bunch. By contrast, overall backgrounds in LER-VACS are driven by the exceptionally large regular backgrounds caused by vacuum scrubbing.

For all injection study data sets, the beam currents and thus the regular beam backgrounds are relatively low. For three of the four HER data sets of the study, injection and regular beam background levels are similar and small. As an exception, HER-PS shows significantly elevated levels of injection backgrounds which surpass all other HER data sets. The ratios of $E_{ib}/E$ for all four HER sets are subject to fluctuations and vary between \SI{22}{\percent} and \SI{86}{\percent} which reflects the different levels of injection backgrounds for the different injection settings. By contrast, the four LER injection study sets clearly show overall background levels comparable to the ones for LER-ALL, and are dominated by injection backgrounds. This can be seen in the ratios of $E_{ib}/E$, where the sets show values of \SI{95}{\percent} and above.

Across all injection study data sets, the fraction of $E_{ib}$ which is deposited within the first \SI{500}{\us} is relatively stable for a respective ring.

Compared to the reference injections, variations in the phase shift in both rings (given by HER-PS and LER-PS) are found to have the largest impact on the injection backgrounds. For HER-PS and LER-PS, the $E_{ib}$ are significantly larger than in the other sets for the respective ring. A larger overall background level also increases the probability for later hits which are above the predefined \SI{2}{MIP} limit. As a consequence, the $t(E_{ib}<$ \SI{2}{MIP}) for these two sets with changed phase shift is also longer.

Finally, no significant increase in the energy levels or decay times is observed for variations in the vertical steering (HER-VS1, HER-VS2 and LER-VS). A possible explanation for this finding is that, due to their position in the horizontal plane, the CLAWS sensors are more sensitive to horizontal than to vertical deviations.

\section{Time pattern analysis}
\label{sec:timing}

As discussed in the previous sections, the time structure of injection backgrounds is largely determined by the propagation of the newly injected particles in the accelerator lattice, which is affected by several different recurring time patterns connected to properties of the accelerator. In this section, a comprehensive study of the periods of the different time patterns is performed. It is based on an analysis of the autocorrelation of the CLAWS measurements. 

\subsection{Analysis methodology}
\label{sec:timint:analysis}

The autocorrelation analysis is based on a discrete autocorelation using the averaged time series of backgrounds, as shown in Figure~\ref{fig:time_structure:her}, as input. Here, the magnitude of the autocorrelation for a given time distance $d_{k}$ is given by
\begin{equation}
 R_{AA}(d_k) = \sum_{i} (A_i \times A_{i-k}),
\end{equation}
with the time distance given by the size of one time bin of \SI{0.8}{\ns} multiplied by the shift in the correlation $k$, 
\begin{equation}
 d_k = \SI{0.8}{\ns} \times k,
\end{equation}
and the sum running over the full length of the waveform.

The result of the autocorrelation is a spectrum of distances in time between signals with their corresponding magnitudes. In this spectrum, the distances which are particularly common, or are separating large signals, appear as peaks. These peaks are connected to regular patterns in the time structure of background signals, making these patterns easily accessible for further study. This enables the identification of the underlying accelerator properties that drive the observed patterns.

\subsection{Autocorrelation results}
\label{sec:timint:results}

\begin{figure*}[ht]
\centering
\includegraphics[width=\textwidth]{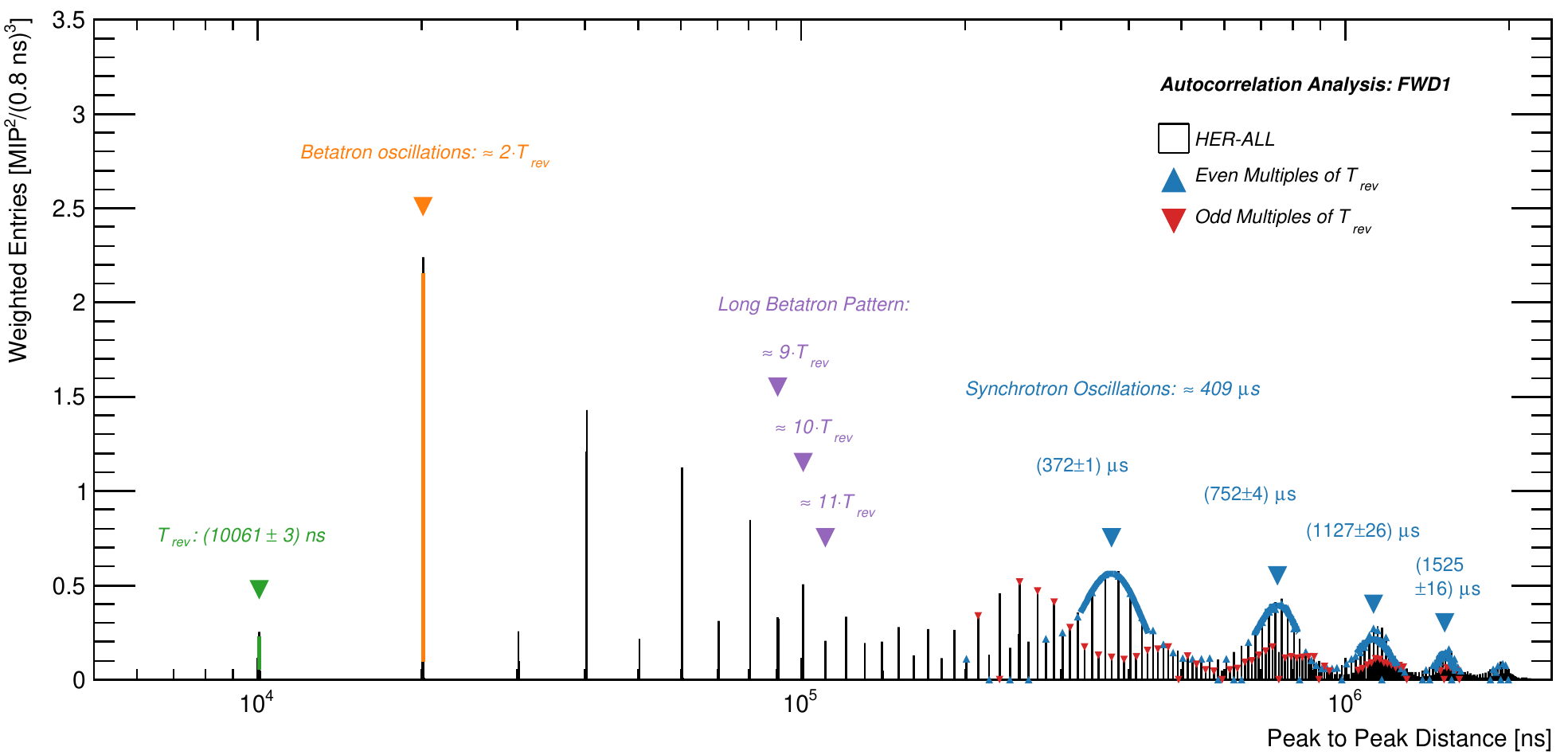}
\caption{Autocorrelation of the averaged background time series for the HER-ALL data set for channel FWD1. Colored Gaussian fits and corresponding labels indicate responses of specific timing patterns connected to properties of the accelerator. For long time differences, even and odd multiples of the revolution time $T_{rev}$ are indicated separately, since the pronounced betatron oscillations, which lead to large differences in amplitude between subsequent revolutions, affect the observable time pattern. See text for details.}
\label{fig:timing}
\end{figure*}

Since the background signals are significantly more extended in time in the HER than in the LER, the HER data provide better access to extended timing patterns. The following discussion thus concentrates in this data set, mentioning LER specific aspects where relevant. Figure~\ref{fig:timing} shows the result of the autocorrelation applied to the HER-ALL data set. On a semi-logarithmic scale in time, the spectrum ranges from time patterns which take place on the time scale of several nanoseconds up to patterns that extend over milliseconds. Very short time scales are dominated by large values originating from neighboring bunches as well as from the temporal extension of individual reconstructed signals over a few nanoseconds. This results in a broad peak extending to around \SI{10}{\ns}, which cannot be resolved for further analysis and is thus not shown in the figure. The first visible structure originates from a single turn of the accelerator at approximately \SI{10}{\us}. 

For events with double bunch injections in the LER, where two bunches simultaneously receive new particles, an additional peak emerges for a time distances of \SI{96.7 \pm 0.2}{\ns}, which is in  agreement with the separation in time for the two injection bunches in that mode of operation of \SI{96.3}{\ns}. This structure is not visible in Figure~\ref{fig:timing} since it shows HER data where only single bunch injection was used. 

In the following, the features of the autocorrelation of the HER-ALL data set is discussed in more detail, with additional remarks on other datasets where relevant differences appear. Quantitative results for identified maxima in the autocorrelation and the corresponding uncertainties are given by the mean of a fit with a Gaussian distribution.\footnote{A two stage maximum likelihood fit in the range of \SI[parse-numbers = false]{\pm 2}{\sigma} of the mean is used; uncertainties on the parameters are adopted from the fit.}

\paragraph{Revolution time}

The first peak in Figure~\ref{fig:timing} is due to the beam revolution period, $T_{rev}$ (\SI{10061.4}{\ns}). Here, we obtain a mean value of \SI{10061 \pm 3}{\ns}, which is, again, in good agreement with the prediction by the machine. Results for all other data sets generally show a similar level of agreement. The rest of the spectrum is composed of multiples of $T_{rev}$ and, in that way, by the propagation of the injected bunches along the beam lines. These timing patterns affect the variation of backgrounds between consecutive transits of the injected bunches.

\paragraph{Betatron oscillations}

For backgrounds recurring with every transit of the injected bunch, the distance corresponding to a full turn in the ring should be the most abundant and therefore the dominant signal in the autocorrelation spectrum. Across all data sets, however, this is not the case. Instead, the largest peak is located at a distance of two consecutive turns (i.e. $2\times T_{rev}$). As discussed in Section~\ref{sec:time_structure:time_patterns}, this is attributed to large amplitude betatron oscillations of the newly injected particles for which the betatron tune has an approximately half integer value. Qualitatively, this behavior appears to be more pronounced for injections into the LER than into the HER, which is consistent with the observed time structure (see Figures~\ref{fig:time_structure:her} and~\ref{fig:time_structure:ler}). This behavior gets further amplified for injections which use a non-optimal phase shift during injections given by the HER-PS and LER-PS sets.

For several data sets, the peaks corresponding to distances from nine to eleven turns are clearly surpassing all but the peak for a distances of two consecutive turns. This suggests that the non-integer part of the vertical betatron tune (see Section~\ref{sec:time_structure:time_patterns}) is deviating from a strict half-integer value by around \num{0.05} oscillations per turn, roughly in agreement with the deviation by \num{0.03} from the half-integer value predicted by simulations. This behavior is generally more pronounced in LER injection data sets than in HER sets, in line with the overall larger betatron oscillation signals in the LER.


\paragraph{Synchrotron oscillations}

From the timing patterns accessible via the autocorrelation analysis, patterns related to syn\-chro\-tron oscillations of the beam particles are the ones occurring on the longest time scales. In the majority of the events recorded by CLAWS, the level of injection backgrounds decays to the level of regular beam backgrounds after \SI{500}{\us} (around 50 turns), as discussed in Section~\ref{sec:decay_behavior}. As such, only a small fraction of the events exhibit synchrotron oscillations beyond more than one period. It should be noted that even if these are only few events, they are the ones which are ultimately determining the length of the gating windows for the rejection of injection backgrounds. Due to the long periods compared to typical background decay times, in the presented analysis synchrotron oscillations can only be observed for data sets with large statistics (i.e. HER-ALL).

Since the pronounced betatron oscillations affect all higher multiples of the beam revolution period in the autocorrelation analysis, two separate distributions are formed for the study of very long oscillation pattern. Beginning with the 20$^{\mathrm{th}}$ multiple of the beam revolution period, the positions of all peaks are determined by a Gaussian fit as mentioned above. The resulting positions of the local maxima are then split into even and odd multiples of $T_{rev}$ as indicated by small blue and red markers in Figure~\ref{fig:timing}. Here, the strong amplification of signals with a period of 2 $T_{rev}$, which manifests as the maximum of the autocorrelation distribution at approximately \SI{20}{\us}, results in stronger signals for even multiples than for odd multiples over most of the time range of the distribution. 

The even multiples reveal a structure of several normal-distributed peaks. These peaks are again fitted with Gaussians for which the determined mean values are given in the figure.\footnote{A maximum likelihood fit with an arbitrary range us used, the uncertainties on the mean are taken from the fit.}
The value of the first mean, \SI{372 \pm 1}{\us}, represents a concrete measurement of the period of the synchrotron oscillations of particles in the HER; the other mean values are multiples of the synchrotron oscillation period. This result is in approximate agreement with the simulation prediction of an oscillation period of \SI{409}{\us} (40.6 turns, see Section~\ref{sec:time_structure:time_patterns}). This finding, however, must be interpreted with some caution since the studied data set covers a wide range of accelerator conditions.

Injection backgrounds in the LER show particle loss rates which are considerably higher early on and are more localized in time, with very little activity beyond \SI{500}{\us} after the injection. As a result, long lasting effects like synchrotron oscillations are less pronounced in the data sets. A decisive measurement of the synchrotron oscillation period with the same methodology as used for the LER-ALL data set as described above is not possible, since most events show at most one synchroton-oscillation related signal enhancement of the injection background. A measurement of this parameter for the LER is thus not reported here. 

\section{Summary and Outlook}
\label{sec:summary}

The CLAWS detector systems has performed extensive studies of beam backgrounds during the first phase of the commissioning of SuperKEKB, referred to as Phase~1.
The sub-nanosecond time resolution of the sensors and the low threshold and good energy resolution which enables the detection of single particles allow time-resolved bunch-by-bunch measurements of beam-induced and injection backgrounds.

The time structure observed during injections shows clear peaks of significantly elevated background rates during the recurring transits of injection bunches, as well as significantly lower, uniformly distributed beam-induced back\-grounds. The comparison of the hit energy distributions of regular and injection backgrounds demonstrates that the signal of regular backgrounds originate from the detection of single charged particles, while the large signals of the injection background are caused by the simultaneous detection of many particles. Nevertheless, the hit energy distributions also show that instantaneous background levels observed during injections in Phase~1 did not exceed regular beam backgrounds by more than one order of magnitude.

Injections into the LER generally result in significantly higher background levels than those into the HER.
This is due to the absence of the positron damping ring, which was only taken into operation after Phase 1.
For both rings, the majority of the injection backgrounds are normally observed within the first \SI{500}{\us} after the injection. The individual decay times back to the regular background levels are fluctuating strongly from injection to injection, and also depend on the exact injection parameters used, in particular on the phase shift.

The different timing patterns of the injection backgrounds are studied with an autocorrelation analysis. The results show effects of betatron and synchrotron oscillations of the particles in the accelerator rings, which influence the background rates over extended time periods. 

Together, the presented studies give insights into the beam backgrounds of SuperKEKB, in particular the background originating from the continuous top-up injection required to achieve high luminosities. The time structure and the measurement of the duration of the elevated background levels provide important information for the further operation of the Belle~II experiment. 

The large fluctuations observed in injection background behaviour and the significant changes in the accelerator and in its operational parameters for physics operation underline the importance of studies and monitoring of injection and other backgrounds also in later phases of SuperKEKB. Based on the successful use of the CLAWS system in Phase~1 reported here, CLAWS was also used in Phase 2 in a different layout, with two ladders with eight channels each as part of the modified inner detector of Belle~II. This system performed studies on injection backgrounds during Phase~2 with even longer continuous monitoring as the one reported here. The results will be reported in an upcoming publication. For the physics runs in Phase~3, a total of 32 further developed CLAWS sensors are installed on the cryostat of the final focusing system (QCS). The injection background-induced particle rates in this location are typically too low for a detailed study. Instead, the CLAWS sensors are now used to monitor large background spikes due to unstable beam conditions, and provide a fast beam abort signal in case dangerous background levels are detected. Further uses for CLAWS sensors as background rate monitors in collimator locations around the SuperKEKB ring are being investigated at the time of writing.

\begin{acknowledgements}
The authors thank Yong Liu, formerly University Mainz, for the scintillator material used for the fabrication of the scintillator tiles. The work was partially supported by the DFG Excellence Cluster ‘‘Origin and Structure of the Universe’’ of Germany and by the European Union's Horizon 2020 Research and Innovation programme under Grant Agreement no. 654168.
\end{acknowledgements}

\bibliographystyle{spphys_TextDOI}

\bibliography{CLAWS_Phase1}

\begin{thebibliography}{10}
\providecommand{\url}[1]{{#1}}
\providecommand{\urlprefix}{URL }
\providecommand{\doi}[1]{DOI: #1}
\providecommand{\eprint}[1]{{#1}}
\providecommand{\archivePrefix}{arXiv}
\providecommand{\primaryClass}[1]{{#1}}

\bibitem{Akai:2018mbz}
K.~Akai, K.~Furukawa, H.~Koiso, Nucl. Instrum. Meth. \textbf{A907}, 188 (2018).
\newblock \doi{10.1016/j.nima.2018.08.017}

\bibitem{Bona:2007qt}
M.~Bona, et~al.,   (2007),  arXiv:0709.0451 [hep-ex]

\bibitem{Abe:2010gxa}
T.~Abe, et~al.,   (2010),  arXiv:1011.0352 [physics.ins-det]

\bibitem{Lewis:2018ayu}
P.M. Lewis, et~al., Nucl. Instrum. Meth. \textbf{A914}, 69 (2019).
\newblock \doi{10.1016/j.nima.2018.05.071}

\bibitem{Gabriel2019}
M.~Gabriel, {CLAWS} - a novel time resolved study of backgrounds during the
  first commissioning phase of {SuperKEKB}.
\newblock Dissertation, Technische Universität München, München (2019).
\newblock
  \urlprefix\url{https://publications.mppmu.mpg.de/2019/MPP-2019-338/FullText.pdf}

\bibitem{firstbeams}
CERN Courier \textbf{56}(3), 11 (2016).
\newblock \urlprefix\url{https://cds.cern.ch/record/2139891}

\bibitem{Piwinski:1998qs}
A.~Piwinski,   (1998),  arXiv:physics/9903034 [physics.acc-ph]

\bibitem{Mori:2014lda}
T.~Mori, N.~Iida, M.~Kikuchi, T.~Mimashi, Y.~Sakamoto, S.~Takasaki, M.~Tawada,
  in \emph{{Proceedings, 5th International Particle Accelerator Conference
  (IPAC 2014): Dresden, Germany, June 15-20, 2014}} (2014), p. MOPRO025.
\newblock \doi{10.18429/JACoW-IPAC2014-MOPRO025}

\bibitem{Hahn:1404985}
F.~Hahn, F.~Ambrosino, A.~Ceccucci, H.~Danielsson, N.~Doble, F.~Fantechi,
  A.~Kluge, C.~Lazzeroni, M.~Lenti, G.~Ruggiero, M.~Sozzi, P.~Valente,
  R.~Wanke, {NA62: Technical Design Document}.
\newblock Tech. Rep. NA62-10-07, CERN, Geneva (2010).
\newblock \urlprefix\url{https://cds.cern.ch/record/1404985}

\bibitem{Sefkow:2018rhp}
F.~Sefkow, F.~Simon, J. Phys. Conf. Ser. \textbf{1162}(1), 012012 (2019).
\newblock \doi{10.1088/1742-6596/1162/1/012012}

\bibitem{Simon:2010hf}
F.~Simon, C.~Soldner, Nucl.Instrum.Meth. \textbf{A620}, 196 (2010).
\newblock \doi{10.1016/j.nima.2010.03.142}

\bibitem{Liu:2015cpe}
Y.~Liu, V.~Büscher, J.~Caudron, P.~Chau, S.~Krause, L.~Masetti, U.~Schäfer,
  R.~Spreckels, S.~Tapprogge, R.~Wanke, in \emph{{Proceedings, 21st Symposium
  on Room-Temperature Semiconductor X-ray and Gamma-ray Detectors (RTSD 2014):
  Seattle, WA, USA, November 8-15, 2014}} (2016), p. 7431118.
\newblock \doi{10.1109/NSSMIC.2014.7431118}

\bibitem{mppc:2019}
{Hamamatsu Photonics}, \emph{MPPC S13360-1325PE}.
\newblock
  \urlprefix\url{https://www.hamamatsu.com/eu/en/product/type/S13360-1325PE/index.html}

\bibitem{Simon:2013zya}
F.~Simon, C.~Soldner, L.~Weuste, JINST \textbf{8}, P12001 (2013).
\newblock \doi{10.1088/1748-0221/8/12/P12001}

\bibitem{Adloff:2014rya}
C.~Adloff, et~al., JINST \textbf{9}, P07022 (2014).
\newblock \doi{10.1088/1748-0221/9/07/P07022}

\bibitem{pico6404d:2018}
{Pico Technology}, \emph{PicoScope 6000 Series}.
\newblock
  \urlprefix\url{https://www.picotech.com/oscilloscope/6000/picoscope-6000-overview}

\bibitem{Ohnishi:2016yyh}
Y.~Ohnishi, et~al., in \emph{{Proceedings, 2nd North American Particle
  Accelerator Conference (NAPAC2016): Chicago, Illinois, USA, October 9-14,
  2016}} (2017), p. MOB3IO01.
\newblock \doi{10.18429/JACoW-NAPAC2016-MOB3IO01}

\end{thebibliography}

\end{document}